\documentclass[reqno, 12pt, a4paper]{amsart}
\usepackage[dvipsnames]{xcolor}
\usepackage{tikz-cd}
\usepackage[utf8]{inputenc}
\usepackage[T1]{fontenc}
\usepackage{amsmath}
\usepackage{amsfonts,amssymb}

\usepackage{tabularx}

\def\[#1\]{%
  \begin{align}#1%
  \end{align}%
}

\definecolor{azulESI}{HTML}{1266AE}
\definecolor{limegreen}{HTML}{32cd32}
\newcommand{\code}{\href{https://doi.org/10.5281/zenodo.19236121}{\raisebox{-.12\height}{\includegraphics[height=1.6ex]{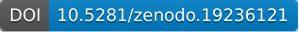}}}}
\usepackage{float}
\usepackage{subcaption}
\usepackage{amsthm}
\numberwithin{equation}{section}
\newtheoremstyle{mytheoremstyle} 
    {10pt}                    
    {8pt}                    
    {\itshape}                   
    {}                           
    {\scshape}                   
    {.}                          
    {.5em}                       
    {}  

\usepackage{hyperref}
\hypersetup{
  colorlinks   = true, 
  urlcolor     = gray, 
  linkcolor    = RoyalPurple!90!black, 
  citecolor   = darkgray 
}

\newcommand{\ja}{\raisebox{.2\height}{\includegraphics[height=1ex]{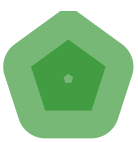}}}
\newcommand{\nein}{\raisebox{.2\height}{\includegraphics[height=1ex]{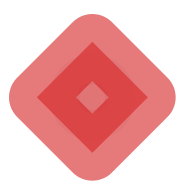}}}
\newcommand{\kreuz}{\raisebox{-.10\height}{\includegraphics[height=1.5ex]{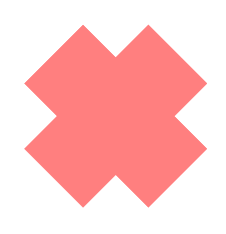}}}
\makeatletter
\newcommand{\leqnomode}{\tagsleft@true}
\newcommand{\reqnomode}{\tagsleft@false}
\makeatother
\theoremstyle{mytheoremstyle}

\newtheorem{theorem}{Theorem}[section]

\newtheoremstyle{definition} 
    {8pt}                    
    {5pt}                    
    {}                   
    {}                           
    {\scshape}                   
    {.}                          
    {.5em}                       
    {}  

 \theoremstyle{definition}

\newtheorem{example}[theorem]{Example}

\newcommand{\inv}{^{-1}}

\newcommand{\Z}{\mathbb{Z}}
\newcommand{\E}{\mathbb{E}}

\newcommand{\C}{\mathbb{C}}
\newcommand{\R}{\mathbb{R}}
\newcommand{\dif}{{\mathrm{d}}}

\newcommand{\M}[1]{M_{#1}(\mathbb{C})}

\renewcommand{\H}{\mathrm{H}}

\newcommand{\ee}{\mathrm{e}}

\newcommand{\MC}{\texttt{MC}}
\newcommand{\True}{\texttt{True}}
\newcommand{\False}{\texttt{False}}

\DeclareMathOperator{\diver}{\mathrm{div}}

\DeclareMathOperator{\Tr}{Tr}

\DeclareMathOperator{\tr}{\mathrm{tr}}

\newcommand{\hp}[1]{^{(#1)}}

\usepackage[includeheadfoot,margin=2.5cm]{geometry}
\usepackage[font=small,labelfont=bf,tableposition=top]{caption}
\usepackage{multicol,caption}

\tikzcdset{arrow style=tikz, diagrams={>={Stealth[round,length=4pt,width=4.95pt,inset=2.75pt]}}}

 \title[Critical curve of  $q$-deformed two-matrix models, Part I]{Critical curve of two-matrix models  $ABBA$, \\ $A\{B,A\}B$ and $ABAB$, Part I: Monte Carlo}
 \author[   C. I. P\'erez S\'anchez   ]{Carlos I. P\'erez S\'anchez }

  \address{ }
  \email{}

\usepackage{stackrel}

\makeatletter

\newcommand*\notocchapter[1]{%
  \if@openright\cleardoublepage\else\clearpage\fi
  \thispagestyle{empty}\global\@topnum\z@
  \@afterindenttrue
  \let\@secnumber\@empty
  \@makeschapterhead{#1}\@afterheading
}

\makeatother

\begin{document}
\begin{abstract} For a family of two-matrix  models
\[
\frac{1}{2} \Tr (A^2+B^2) - \frac{g}{4}  \Tr (A^4+B^4)
- \begin{cases}
\frac{h}{2}  \Tr (  A BA B)     \\
\frac{h}{4}  \Tr (  A BA B+ ABBA )    \\  \frac{h}{2}  \Tr (  A B BA )
  \end{cases} \notag
 \]
with hermitian $A$ and $B$, we provide, in each case, a Monte Carlo estimate of the boundary of the maximal convergence domain in the $(h,g)$-plane. The results are discussed comparing with exact solutions (in
agreement with the only analytically solved case)
and phase diagrams obtained by means of
the functional renormalisation group.
\end{abstract}
 \maketitle%
\vspace{-2ex}
\begin{figure}[h!]
\includegraphics[width=.75\textwidth]{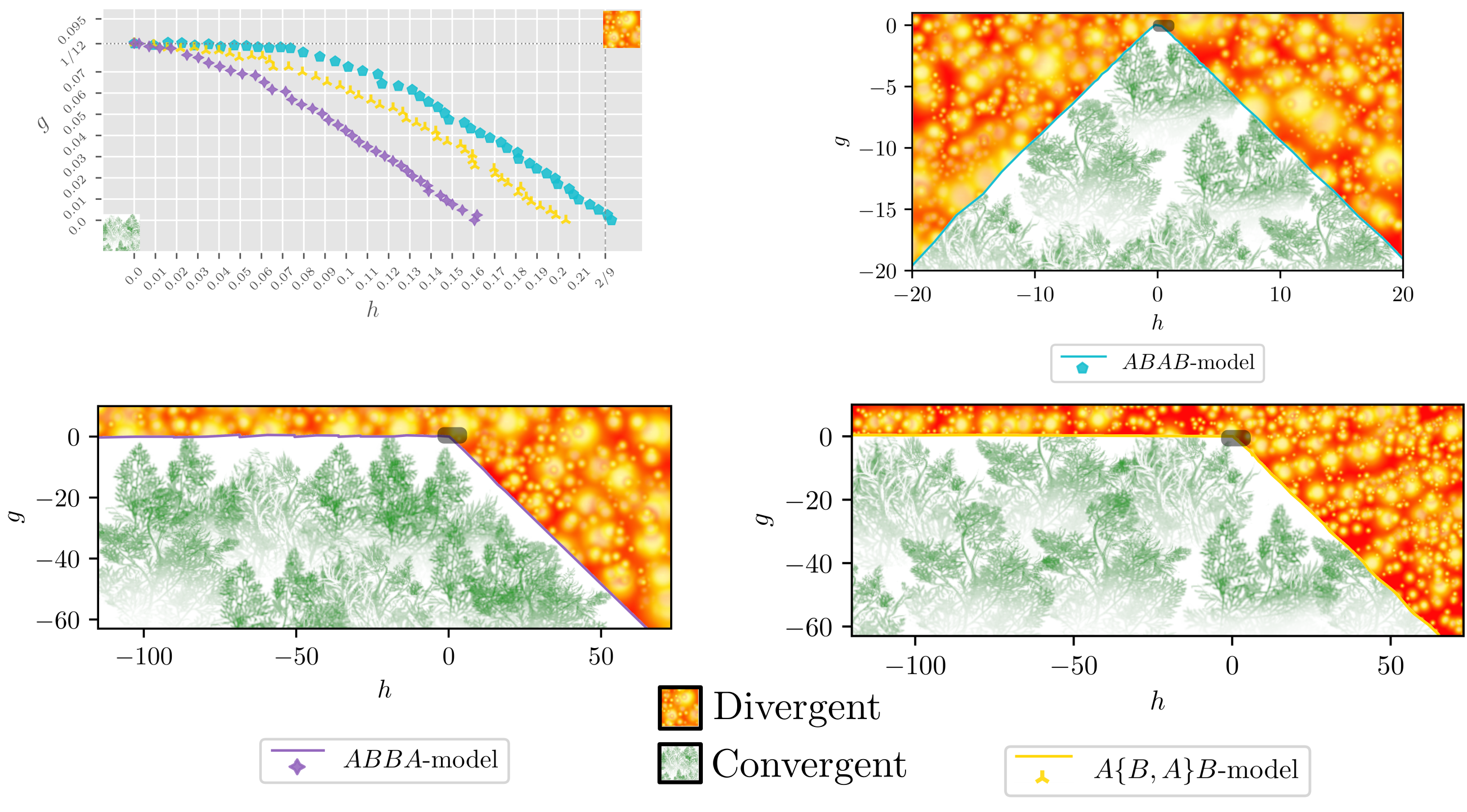}
\caption{Sketch of the phase diagram of  three models.
The top left plot lies inside the three darkened regions
of each of the  three other plots.
\label{fig:intro}}
\end{figure}

\allowdisplaybreaks[3]
\section{Introduction}\label{sec:Intro}
We determine the critical loci of a family of integrals over  two
hermitian random matrices, $A $ and $B$, of a large size $N$.
For $q\in  [0,1]$, $g,h\in \R$, let
\[ \label{action}
S_{g,h} \hp q(A,B) =\frac{1}{2} \Tr (A^2+B^2) - \frac{g}{4}  \Tr (A^4+B^4)
- \frac{h}{2}  \Tr (  A \{B,A\}_q B)
\]
with the following notation
\[ \{B,A\}_q:=qBA + (1-q)AB \text{, with }  0 \leq q \leq 1 .  \label{acomm}\]
It is part of our present
work to numerically determine the maximal real domains of the couplings  $g$ and $h$  for which
the partition function
\[ \label{intro_ZN}
Z\hp{q} _N (g,h) =  \int \ee^ { - N S_{g,h} \hp q(A,B) } \dif A\,\dif B
\]
exists in the cases $q=0, \tfrac 12 $ and $1$. (The integration for each matrix is performed over its $N^2$ real degrees of freedom and the measure is so normalised, that at $g=h=0$ the partition function is $1$, see Eq. \eqref{dAdBmeasure} for details.) Setting $q=1$ gives rise to the  $ABAB$-model (we will name models
 using essentially the operator accompanying $h$), which was
analytically solved by Kazakov--Zinn-Justin (Jr.) in \cite{ABAB}. The $ABAB$-model
 is special, for one because
$\Tr ABAB$ is not a positive-definite operator (this, as we will see, makes the
phase diagram   non-trivial) that, unlike $\Tr A^3$, does mix the matrices. But the $ABAB$-model is also important because
it was solved by the character expansion of \cite{KSW},
which proved fruitful to others (e.g. \cite{Benedetti:2008hc}) and does so  even at present \cite{Abranches:2023tqe}.\\

Unlike one-matrix models,  two-matrix models can be solved only in exceptional cases.  The full solution of one-matrix models was streamlined by Eynard-Orantin in  their topological recursion \cite{TR} (if needed, to gain familiarity, see \cite{EynardCounting,BorotNotes} before);
a family of two-matrix models can be solved by the same tool \cite[Ch. 8]{EynardCounting}, but for an arbitrary potential a general analytic tool is not available. This happens with the models \eqref{action}.
It would be interesting to study these using topological recursion or the character expansion\footnote{The character expansion
is very unlikely to help here, since in the Kazakov--Zinn-Justin solution of the $ABAB$-model,
that method relied on the companion operator of $h$ being  (the trace of) an analytic function  $F(z)=z^2$ of $AB$;
for $q<1$, the analogue function  $F_q(z)= q z^2 +(1-q) | z|^2$ depends on the conjugate
of $z$. }, but for the time being we resort to computer  simulations. \\

Indeed, when $h=0$, the potential does no longer mix $A$ with $B$ and the two-matrix model will
factor as  $Z\hp q_N(g,0) = [z_N(g)]^2$, where $z_N(g)=\int \ee^{-N \Tr  M^2/2 + Ng \Tr M^4/4 }\dif M$.
This model is critical at $g=+ 1/12$
in the present  sign convention. Thus, for each $q$ in \eqref{intro_ZN},
the critical point $(g,h)  = (1/12, 0)$  of pure gravity lies on the \textit{critical curve}, the
boundary of the maximal region on the $(g,h)$-plane where the model exist.
Our Monte Carlo simulations find such critical curve and the asymptotics for large $|h|$.
The first  coarse conjecture---after a small excursion to the $q=\tfrac 34$-model that supports it---is that, \textit{in that limit}, Models \eqref{action} come in two flavours:
\[\qquad\raisebox{-2ex}{\includegraphics[width=.5\textwidth]{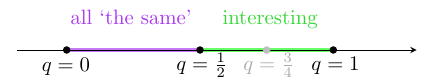}} \label{q-line}\]
When $q=1$, the $h$-operator $\Tr ABAB$ is oscillatory,
as it is non-trivially integrated under the U$(N)$-relative angle between the two matrices that is present
in \eqref{intro_ZN}; on the other end,  $\Tr ABBA=\Tr [AB(AB)^*]$ is constant under unitary integration. The expectation of the $\Tr ABAB$ operator is seen (numerically, see Fig. \ref{fig:sign_ABAB}) to
take the sign of $h$, so the product $ h \Tr ABAB$ has always the same sign in expectation,
regardless of $h$ (if $g$ allows for finiteness). The U$(N)$-relative angle becomes then dominated by a positive part inside the
$h$-operator from $q=\tfrac 12$, yielding `the same' asympotics in $q \in [0,\tfrac 12]$.\\

It is pertinent to mention that Monte Carlo is only a choice to approach this problem. Other numerical techniques emerged more recently in the context of lattice Yang-Mills theory in \cite{Kruczenski}, who built a matrix of moments (that is essentially indexed by
loops in the lattice, and satisfies the Toepliz property) and proved that this matrix should be positively defined. Positive conditions
have been also notably employed in \cite{lin2020bootstraps} for matrix models of several hermitian variables. In both cases, the entries of the matrix of moments can be computed recursively, using
loop or Schwinger-Dyson equations, in terms of a finite family of moments. The positivity bounds
yield a parametrisation of all moments
in terms of the coupling constants, which being so sharp (see in particular \cite{Kazakov:2021lel}), effectively yields a solution of the model, the cut off being now no longer the size of the random matrices
but that of the matrix of moments.
In each case, the proof of positivity is itself trivial, but its consequences are powerful and seemingly
enjoy of a universal applicability \cite{lin2020bootstraps,hessam2022bootstrapping,perez2025loop,berenstein2024one,samuel_bootstrap,Khalkhali:2020jzr, Reiko_Bootstrap, NathanCarlosBrayden}.
In the case at hand, even though it
should be possible  to `bootstrap'  the models \eqref{action} for all $q$ inside a single program using this approach, it is advantageous to have Monte Carlo simulations ready as an independent method
to contrast with.
Part II of this paper series addresses the viewpoint of criticality by using  positivity conditions for
bivariate matrix models  like \eqref{action}. \\

Other method to estimate the critical loci of multi-matrix models is the functional renormalisation group
which seems effective
for a first exploration, as it requires less resources than Monte Carlo and deliver a prompt, in case of random matrices, qualitative  answer. In the case of multi-matrix models,  this approach is still in the process of telling apart criticality for different models: in fact, by zooming out the `phase diagram of the $ABAB$-model'  that was
derived in \cite{ABAB_FRG} one observes the lack of the symmetry
 $h\mapsto -h$ commented under \eqref{q-line} and one discovers an asymptotic behavior that, thanks to the present simulations, can be identified as characteristic of a $q$-model with $q\leq \tfrac 12$ (cf. Sec. \ref{sec:bonus} to see why this paragraph is in part also self-critique).\\

The code \cite[$\code$]{perez_sanchez_code} that we designed to perform the experiments of the present article should immediately work
for other bivariate two-matrix models, and after an obvious extension, for
any multi-matrix model. Commencing with the relevant methods for
multi-matrix models in Section \ref{sec:2MM}, we then briefly explain
the Hamiltonian Monte Carlo (MC) approach in  Section \ref{sec:MCMC}. Our code was benefited from the
openness of the code of \cite{JhaMC}\footnote{This reference provides a code written in \texttt{Python}; for \texttt{C++} code, we refer to
e.g. \cite{BarrettGlaser,DArcangelo:2022rfb} who addressed a related problem, if with more structure (and rather using Metropolis-Hastings Monte Carlo).}. We modified part of it to match our needs and extended it by
new methods to find the critical curve. Our implementation is described in Section \ref{sec:Methods}.  Then Section \ref{sec:Results}
 presents the results and Section \ref{sec:conclusions} closes with the perspective and conclusions.
To finish, we remark that all plots before Section \ref{sec:Results}
are actual simulations, but were added only for exposition and notation purposes. However, only the  plots of Section \ref{sec:Results} were taken into account in the final statistics.

\tableofcontents

\section{Two-matrix models}\label{sec:2MM}

Let us denote by $\H_N$  the space of $N\times N$ (complex) hermitian matrices.
We  compute the correlators or expectation values of \textit{words} $W_{A,B}$ in two letters  (monomials $ W_{A,B} \in \C\langle A,B\rangle$ like $A^4BAB$) by
\[
\mathbb E  [ \tfrac 1N \Tr (W_{A,B}) ]  = \frac{1}{Z  } \int _{\H_N\times \H_N} \frac 1N\Tr (W_{A,B})
 \ee^ { - N S (A,B) } \dif A\,\dif B . \label{correlator}
\]
Of course, we should write $\mathbb E \hp q_{g,h; N}$,
$Z\hp{q} _N (g,h) $ and $S_{g,h}\hp q $ instead of $\mathbb E, Z$ and   $S$,
but we use this heavier notation only if the context is not clear.
In Eq. \eqref{correlator} the measure is given by  \[ \label{dAdBmeasure}
\dif M = {2}^{-N/2} {(N/\pi)}^{N^2/2} \prod_{i=1}^N \dif M_{ii} \prod_{j=i+1,\ldots,N } \dif \,\mathrm{Re} M_{ij} \dif \, \mathrm{Im} M_{ij} \qquad (M=A,B )
\]
which yields $\dif A\,\dif B$ normalised in the sense that $Z(0,0)=1$,
and Eq. \eqref{correlator} yielding $1$ for the empty word $W_{A,B}=1_N$  (this also
fixes notation for the trace).\par

\subsection{Positivity} \label{sec:Positivity}
For a fixed matrix model (fix $q$), since
$\sum_{a,b=1} ^ N M_{a,b} \bar M_{a,b} >0 $ holds
for any matrix $M \in M_N(\C)$
so does for  any
complex linear combination
$M= \sum_{i=1}^L z_i W_i $  of arbitrarily many ($0<L<\infty $)
words $W_i \in \C \langle A,B\rangle$ in the letters $A$ and $B$, $z_i \in \C$. Thus
\[ \label{ineq_for_M}
0 < \sum_{i,j=1}^L z_i   \Tr( W_i W_j ^*)  \bar z_j \text {, hence  }
0 < \sum_{i,j=1}^L z_i \E [ \Tr( W_i W_j^* ) ] \bar z_j
\]
for every $z = (z_i)  \in\C^L$. The latter implies the non-negativity of the matrix $\mathbb M $,
\[ \label{positivity_for_M}
\mathbb M \succeq 0\qquad  \text{with entries } \mathbb M _{ij} := \lim_{N\to \infty }\frac{1}{N}\E  [ \Tr W_i W_j ^* ] = \mathbb M _{ji } \,.
\]
By restoring the notation, one sees that $\mathbb M $ of course
depends on $g$ and $h$  and $\mathbb M(g,h) \succeq 0$  restricts the domain
of $(g,h)$. We are already in the large-$N$ regime, so the $N$-dependence disappeared;  in the large-$L$ limit one gets the solution of the model
in the sense of having determined all correlators  \[ \{ \E[W] : W\in \C \langle A,B\rangle, \deg W \leq L \} \] as a numerical estimate in terms of each $(g,h)$, thus also the maximally allowed domain. The
sufficiency of a suitable set of independent parametrising moments
---despite the exponentially growing family---has been addressed by \cite{lin2020bootstraps}.
\par

Although simple, two inequalities will be useful:
\begin{subequations}\label{hierarchies_examples}%
\[
-\Tr ABBA \leq \Tr ABAB  &\leq \Tr ABBA  \quad \text{and }\quad \label{hierarchy_a} \\    \Tr  ABBA  &\leq  \frac 12 \Tr (A^4 +B^4)  \label{hierarchy_b}   .  \]%
\end{subequations}%
 The first hierarchy of operators can be obtained (in particular) by using  Ineqs. \eqref{ineq_for_M} with
 $M = z_1 W_1 + z_2  W_2 $; then the matrix $\Tr (W_iW_j^*)$  is non-negative, hence
$\det  [   (  \Tr W_iW_j^*)_{i,j} ] \geq 0$ which is equivalent to Ineq. \eqref{hierarchy_a}, when $W_1=AB, W_2=BA$. The second one rephrases the trivial fact that
$ 0 \leq \Tr C^* C$ for any square matrix $C$, and especially when $C=A^2-B^2$.
\\

From Ineq. \eqref{hierarchy_a},
which is still deterministic, observe that for fixed couplings $(g,h) $ and matrices $A,B \in \H_N$,
the function $q\mapsto S_{g,h} \hp q(A,B)   $ is non-decreasing in the parameter $q$,
hence the integrand in Eq. \eqref{intro_ZN} is non-increasing  with $q$:
\[
\ee^{-N S\hp 0_{g,h}(A,B)}  \geq
\ee^{-N S\hp {q}_{g,h}(A,B)}  \geq
\ee^{-N S\hp {\tilde q}_{g,h}(A,B)} \geq
\ee^{-N S\hp 1
_{g,h}(A,B)} ,\qquad (g,h) \in \R^2_{\geq 0},  q<\tilde q. \label{ineqs_Sq}
\]
For fixed $g>0$, if there exist a point $(g,h_q(g)  ) $ in the critical curve of the $q$-model, the previous inequality
means that if $(g,h_{\tilde q} (g))$ is in the   critical curve of the $\tilde q$-model, then $h_{\tilde  q}(g) \geq h_{ q}(g)  $.
Restricting ourselves temporarily to $\R^2_{\geq 0}$, we conclude that no point of the
segment of the critical curve for the $q$-model in that positive quadrant
can lie outside the convergent region of the $\tilde q$-model, if $q<\tilde q \leq 1$.
After Section \ref{sec:Methods}, it will be clear, that one saves also \textsc{cpu}-time by starting with the $ABAB$-model, determining its critical curve, and use this information to simulate for models with $q<1$.
\\

In the Introduction we commented on the bound $g \leq 1/12$ (when $h=0$) required
for the models to exist. Again by \eqref{hierarchy_a}, if  $q \in [0, \tfrac 12 ]  $ then $\Tr A\{B,A\}_q B \geq 0$, hence also for $h>0$ one obtains the bound $g \leq 1/12$.  Now we refer to
the exact solutions when $g=0$, which yield a critical $h$ value given by  $h_{q=1}(g=0) = 2/9$
\cite[Sec. 5.3, in terms of the inverse coupling or  temperature, $T_*=4.5$]{Kristjansen} or \cite[App. A]{ABAB} for the $ABAB$-model ($q=1$-case). But then Bounds \eqref{ineqs_Sq}
imply that \[ h_{q'}(g=0) \leq h_{q=1}(g=0) = 2/9 \text{ for any $q'\leq 1$}. \label{hbound} \]
Hence, the segment in $\R^2_{\geq 0}$ of the critical curve
lies in the rectangle $[0,1/12] \times [0,2/9]$ for $q \in [0, \tfrac 12 ] $ and  Ineq. \eqref{hbound}
holds for any $q$. (This is however what we know before simulations. After these,
one observes that also for $q=1$ it holds that $g\leq 1/12$ whenever $h>0$,
since, for any $h$,  $\mathbb E_{g,h} \hp {q=1}  [ \Tr ABAB] $ takes the same sign of $h$, see  Fig. \ref{fig:sign_ABAB}.)

Positivity is also mentioned because its application to solve
path integrals in general is relatively new.
For instance, few is known about the precise value of determinants of (the minors of) $\mathbb M$. Our Python program monitors this. In order to compute the entries of $\mathbb M$,
the loop or Schwinger-Dyson equations, which we address below, are used.

\subsection{Schwinger-Dyson Equations}\label{sec:SDEs}
While it is possible to geometrically describe the origin of the Schwinger-Dyson Equations (cf. \cite{Azarfar:2019nlk}), we here provide the shortest derivation and
focus on them as a tool.  Given two words $ X(A,B),Y(A,B)$ in the two matrix variables $A,B$,
 the field $F = (X,Y) : \H_N\times \H_N \to \H_N\times \H_N $ is smooth.
From   $\int \dif ( F  \ee^{-N S})  =0  $  it follows
\[\int(\diver  F ) \ee^{-N S} = N \int   \sum_{\alpha=1,\ldots, 2 N^2 }  \bigg( F_{\alpha} \partial_\alpha S   \ee^{-NS}   \bigg )\dif A\, \dif B\]
upon identification of $\dif$ with the divergence, as it acts on the vector field $F:  \R^{2\times N\times N} \to  \R^{2\times N\times N}$.
Operationally, it is convenient to pass from $2N^2$ real variables to  equivalent
`coordinate' free definitions that allow to compute only by moving letters, \cite{GuionnetFreeAn,Perez-Sanchez:2020kgq}.
This is not only an advantage to obtain compact expressions for the Schwinger-Dyson
Equations but the code profits from these too while computing the force that
leapfrog integration needs (cf. Sec. \ref{sec:proposals}).
The first half of values for $\alpha$ correspond to $(\partial_A)_{i,j}$ and
the other to $(\partial_B)_{i,j}$ for $i,j=1,\ldots,N$, so one obtains
$\E [ \diver F ] = N \E [ \Tr (X  \partial_A S+ Y \partial_B S)  ]$;
here the expression $X  \partial_A S$  means matrix multiplication of $X$ with $\partial_A S$. In order to
obtain  the matrix derivative $\partial_A$, one defines for $ M_k\in \M N$ and $A\in \H_N$,
\[ \label{cyclicgradient}
\mathsf {D}_A  (M_1 M_2\cdots M_p) &:=
\partial_A \Tr ( M_1 M_2 \cdots M_p)  \\
&:=  \sum_{\substack{ i =1,\ldots,p \\  \text{ if } M_i=A }}
M_{i+1} M_{i+2} \cdots M_p M_1\cdots M_{i-1}.  \notag
 \qquad \quad
\]
The operator $\mathsf {D}_A $ is called cyclic gradient \cite{Voi_gradient}; notice that  $\mathsf D_A W $ is independent of the cyclic
reodrerings of $W$ (and being the derivative of a trace, it must be). Finally,
the divergence of $F=(X,Y)$ is given by $\diver F = \partial_A X+\partial_B Y$. In order
to make sense of this expression one uses the matrix derivatives (or noncommutative
derivatives), which are defined by (here $\delta_{M_k,A}=1$ only if $M_k=A$,
and else $\delta_{M_k,A}=0$, and notice the absence of the trace in the argument)
\[ \label{ncderv} \partial _A  (M_1 M_2\cdots M_p) & := \delta_{M_1,A} 1\otimes M_2\ldots M_p \\
& +
 \sum_{  1< i  < p  } \delta_{M_i,A}\notag
 M_1M_2 \cdots M_{i-1} \otimes M_{i+1}
 \cdots M_{p} \\ & +  \delta_{M_p,A} M_1M_2\ldots M_{p-1} \otimes 1 \notag .
\]
\begin{example} For hermitian $A,B$,
\[ \notag
\mathsf {D}_A (ABBA) & =  ABB  + BBA \\\notag
\mathsf {D}_B  (A^3 BAB^2) &=  AB^2A^3 +  BA^3BA+A^3BAB \\\notag
\partial _A (ABAB) &= 1\otimes BAB + AB \otimes B \\\notag
\partial _B (ABAB) &= A \otimes AB  + ABA \otimes 1.
\]
\end{example}
The Schwinger-Dyson equations
$\E [ \diver F ] = N \E [ \Tr (X  \partial_A S+ Y \partial_B S)  ]$, when  rewritten in the previous notation,
take the neat form (if $S=\Tr V$ with $V(0,0)=0$)
\[ \label{SDEs}
\E   \bigg[ \Big(\frac 1N \Tr \otimes \frac 1N \Tr \Big) ( \partial_A X + \partial_B Y) \bigg]
=\E \big [  \Tr (X\mathsf D_A V + Y \mathsf D_B V) \big].
\]
Using the normalised trace  $\tr = \tfrac 1N \Tr $ is also common\footnote{Caveat:
Another common convention is precisely the opposite (lowercase means there unnormalised and capitalised normalised). For us $\Tr 1=N$. }.
Then explicitly the SDEs read as in Table \ref{tab:SDEs}.
To simplify that list, observe that, for any $q$, the model \eqref{action} has dihedral
symmetry generated by the exchange $(A,B) \mapsto (B,A)$
of the two matrices, and by the sign change of $M\mapsto -M$ for $M=A,B$.
Then the correlator $ \mathbb E [ \Tr (W_{A,B}) ]   $ is non-zero only if
the degrees in $A$ and $B$ are even, $\deg  W_{A,1}, \deg W_{1,B} \in 2\Z_{\geq0 }. $\\

We close this section with notation (or, due to the tacit dependence on $g,h,N,q$, abuse thereof):
\begin{subequations}\label{moments}%
\[
t_2 &= \frac 1{2N} \E [ \Tr (A^2 +B^2)], \quad &&& t_{4} &=  \frac 1{2N} \E [ \Tr (A^4 +B^4)], \\
t_{2,2} &=  \frac 1{N} \E [ \Tr (ABBA)] && &
t_{1,1,1,1}& =  \frac 1{N} \E [ \Tr (ABAB)].
\]
\end{subequations}
The $1/2$ factor should not cause trouble, since due to the dihedral symmetry, the
results turned out to be indeed symmetric under the exchange of the matrices (we
do not plot  $\frac 1{N} \E [ \Tr (A^2 )]$ and $ \frac 1{N} \E [ \Tr (B^2)]$ independently).

\begin{table}
\footnotesize
\begin{align*} X \quad\qquad  &  \qquad\text{Schwinger-Dyson Equation} \\ \hline & \\[-1ex]
A  \text{ yields: }& \tr^{\otimes 2} \big[ {1} \otimes {1} \big] \\  & =
\tr \big[ A^{2} -g A^{4}  -  \, h {(1 - q)} A^{2}B^{2} -h q  A B A B  \big] \\[1ex]
 A^3  \text{ yields: }& \tr^{\otimes 2} \big[ {1} \otimes {A^{2}} + {A} \otimes {A} + {A^{2}} \otimes {1} \big] \\  & = \tr \big[ A^{4} -g A^{6}  - h (1-q) \color{black} A^{4}B^{2} -h q  A^{3}B A B   \big] \\[1ex]
B^2A  \text{ yields: }& \tr^{\otimes 2} \big[ {B^{2}} \otimes {1} \big] \\  & = \tr \big[ B^{2}A^{2} -g B^{2}A^{4}   -  \tfrac12 h (1-q) \color{black} B^{6}A^{2} \\ & \quad -h q  B^{3}A B A    - \tfrac 12 h (1-q) \color{black} B^{2}A B^{2}A \big] \\[1ex] BAB  \text{ yields: }& \tr^{\otimes 2} \big[ {B} \otimes {B} \big] \\  & = \tr \big[  A B A B -g A^{3} B A B    -h q  B A B^{2}A B  - h (1-q) \color{black} B A B^{3}A \big] \\[1ex]
A^5  \text{ yields: }& \tr^{\otimes 2} \big[ {1} \otimes {A^{4}} + {A} \otimes {A^{3}} + {A^{2}} \otimes {A^{2}} + {A^{3}} \otimes {A} + {A^{4}} \otimes {1} \big] \\  & = \tr \big[ A^{6} -g A^{8}  - h (1-q) \color{black} A^{6}B^{2} -h q  A^{5}B A B   \big] \\[1ex]
 B^4A  \text{ yields: }& \tr^{\otimes 2} \big[ {B^{4}} \otimes {1} \big] \\  & = \tr \big[ B^{4}A^{2} -g B^{4}A^{4}  - \tfrac12  h (1-q) \color{black} B^{6}A^{2}\\ & \quad -h q  B^{4}A B A B  -\tfrac12 h (1-q) \color{black} B^{4}A B^{2}A \big] \\[1ex]
 A^3B^2  \text{ yields: }& \tr^{\otimes 2} \big[ {1} \otimes {A^{2}B^{2}} + {A} \otimes {A B^{2}} + {A^{2}} \otimes {B^{2}} \big] \\  & = \tr \big[ A^{4}B^{2}  -g A^{6}B^{2}  - \tfrac12 h (1-q) \color{black} A^{3}B^{2}A B^{2} \\ & \quad -h q  A^{3}B^{3}A B -\tfrac 12 h (1-q) \color{black} A^{4}B^{4} \big] \\[1ex]
 B^3AB  \text{ yields: }& \tr^{\otimes 2} \big[ {B^{3}} \otimes {B} \big] \\  & = \tr \big[ B^{3}A B A -g B^{3}A B A^{3} - \tfrac 12 h (1-q) \color{black} B^{5}A B A \\ &\quad   -h q  B^{3}A B^{2}A B - \tfrac 12 h (1-q) \color{black} B^{3}A B^{3}A \big] \\[1ex]
 A^2B^2A  \text{ yields: }& \tr^{\otimes 2} \big[ {1} \otimes {A B^{2}A} + {A} \otimes {B^{2}A} + {A^{2}B^{2}} \otimes {1} \big] \\  & = \tr \big[ A^{2}B^{2}A^{2} -g A^{2}B^{2}A^{4} - \tfrac 12 h (1-q) \color{black} A^{2}B^{2}A^{2}B^{2}\\ &\quad  -h q  A^{2}B^{2}A B A B - \tfrac 12 h (1-q) \color{black} A^{2}B^{2}A B^{2}A \big] \\[1ex]
 A^2BAB  \text{ yields: }& \tr^{\otimes 2} \big[ {1} \otimes {A B A B} + {A} \otimes {B A B} + {A^{2}B} \otimes {B} \big] \\  & = \tr \big[ A^{2}B A B A -g A^{2}B A B A^{3} - \tfrac 12 h (1-q) \color{black} A^{2}B A B A B^{2}\\ &\quad  -h q  A^{2}B A B^{2}A B - \tfrac 12 h (1-q) \color{black} A^{2}B A B^{3}A \big] \\[1ex]
 ABABA  \text{ yields: }& \tr^{\otimes 2} \big[ {1} \otimes {B A B A} + {A B} \otimes {B A} + {A B A B} \otimes {1} \big] \\  & = \tr \big[ A B A B A^{2} -g A B A B A^{4} - \tfrac 12 h (1-q) \color{black} A B A B A^{2}B^{2}\\ &\quad  -h q  A B A B A B A B - \tfrac 12 h (1-q) \color{black}  B A B A B^{2}A^2 \big]
\end{align*}
\caption{In this table of Schwinger-Dyson Equations for the model \eqref{action},
$X$ is the word listed on the left while $Y=0$ in each column (which we later on in
Fig. \ref{fig:SDE_thermalisation} is referred to as `varying $A$', also see \eqref{SDEs}
for notation). Also $\tr$ is the normalized trace, so $\tr 1=1$, and $\tr^{\otimes 2} ( w_1 \otimes w_1)=\tr w_1 \tr w_2$ for words $w_1,w_2$. Since  Models \eqref{action} are all dihedral symmetric, several terms above vanish, e.g. $\tr^{\otimes 2} ( B\otimes B^3)=0$. \label{tab:SDEs}}
\end{table}

\section{Hamiltonian Markov Chain Monte Carlo} \label{sec:MCMC}
Let  $n$  be a large integer, which in the code will correspond to the number of iterations.  Our aim is to
compute correlators as a sum
\[
\mathbb E  [ \Tr  W ] =\frac 1{n-\tau} \sum_{i=\tau+1 }^{n} \Tr  W(X_i)
\]
up to terms $O( (n-\tau)\inv  )$, where $\tau $ is the thermalisation time
($1< \tau< n$, see Sec. \ref{sec:thermalization} for criteria).  Essentially, the
algorithm that follows will guarantee that, after the thermalisation time $\tau$,  we pick  matrices  $X_i=(A_i,B_i) \in \H_N\times \H_N$  ($i=\tau+1,\ldots, n$)
using the importance sampling, that is
matrices using the  measure \[  \frac{1}{Z}\exp[ - N S(A,B) ] \dif A\,\dif B . \]

The strategy is to use  Hamiltonian  Markov Chain  Monte Carlo (MCMC) simulation. In order to generate the Markov Chain $X_1,X_2,\ldots, X_n$
one fixes criteria to
accept as $X_{i+1}$ a proposal $\tilde X$, once $X_i$ has been accepted in the chain,
and provides a criterion to propose candidates $\tilde X$ to be tested.
The initialisation is $X_1= 0_N $. (In order to save thermalisation time, we  attempted to start from a matrix pair $X_1=(A,B)$ that solves the Schwinger-Dyson equations. With this initial point, several proposals $\tilde X$ were rejected from the very beginning of the simulation; starting from $X_1=0_N$ we achieved better results.)

\noindent
\begin{figure}%
\centering\noindent \hspace{-3ex}%
\begin{minipage}[l]{.7\textwidth}
\centering\captionsetup{width=.7\linewidth}
\includegraphics[width=.7\textwidth]{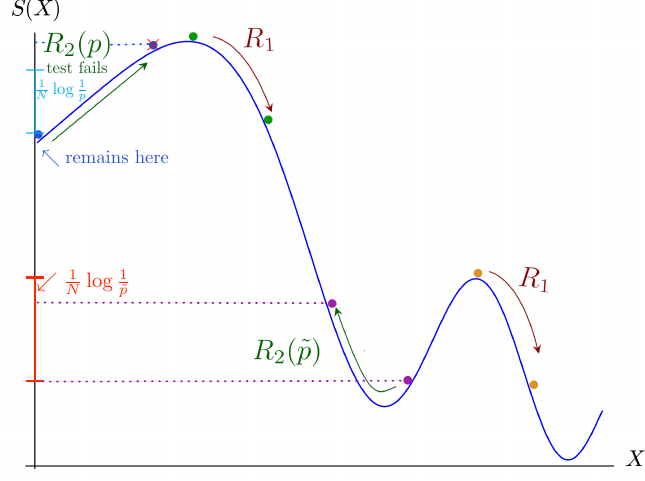}
\caption{Using a fictive potential $S(X)$ (blue curve), rules $R_1$ and $R_2$ are sketched. In each situation, the beads on each arrow tail mean the last Markov
 chain member, and at the arrow tip the proposed move $\tilde X$.  If $\Delta S<0$,  according to $R_1$, one accepts the move.
For $R_2(p)$ in the left-top case, the increment in $S$ leads
to rejection, since $ N \Delta S> \log 1/p$,
 whereas the $R_2(\tilde p)$ is verified in the case of the proposed move for the purple points, as the bound $(1/N) \log (1/\tilde p)$ was not exceeded ($p,\tilde p $ are freshly uniformly chosen
 random numbers in $(0,1)$ for each acceptance test).\label{fig:R1R2}}
    \label{fig:sample_figure}
\end{minipage}\!\!\!\!\!\!\!\!\!\!\!\!\!\!\!\!\!\!\begin{minipage}[c]{.44\textwidth}
\vspace{4ex}\captionsetup{width=.79\linewidth}
\centering\hspace{1ex}{\includegraphics[width=0.9\textwidth]{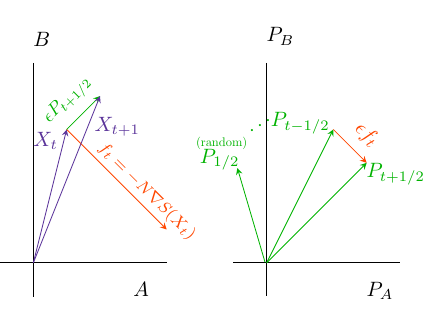}}\vspace{3ex}\caption{On the Hamiltonian approach to MCMC with the leapfrog integrator\label{fig:leapfrog}. The left (right) plane is the `configuration' (resp. momenta) $\H_N^2$-space of matrices $X_t=(A_t,B_t)$ [resp. $P_{t+1/2}=(P_{A,t+1/2},P_{B,t+1/2}) $]. \vspace{3ex}}
    \label{fig:sample_figure}
\end{minipage}
\end{figure}

\subsection{Generation of new proposals}\label{sec:proposals}

We use the Hamiltonian leapfrog integration to generate new configurations $\tilde X$.
More concretely, one duplicates the degrees of freedom,  introducing a second copy
for the momenta space $P= (P_A,P_B) \in \H_{N}^2$, and tackles the equations
\[ \dot P = - \nabla_X H  =- N \nabla_X S, \qquad  \dot X = \nabla_P H = P  \]
for the Hamiltonian given by  $H  = \Tr ( P^2/2)  +  N S $, being the dot
the derivative with respect to a time parameter $t\in [0,T]$.  One calls \textit{force} the matrix field
\[
f := - N \nabla_X S \in \H_N^2, \qquad  \nabla_X = (\partial_A,\partial_B).
\]

The gradient's components are matrix derivatives\footnote{The $(ij)$ entry of a matrix derivative is $\partial/\partial M_{ji}$, which often appears as $\partial_M= \partial/\partial M^T$; we refer to \cite{Perez-Sanchez:2020kgq} for more details, since provide only the expressions needed in the code} defined in Eq. \eqref{ncderv}.
For the potential $S=S\hp q_{g,h}(A,B)$ in Eq. \eqref{action}, the force $f=(f_A,f_B)$ is given by
\begin{subequations}
\[
\frac 1N f_A &=
A   - g A^3 - h q BAB - \frac{1}{2} h (1-q) (BBA + ABB)
\\
\frac 1N  f_B &= B   - g B^3 - h q ABA - \frac{1}{2} h (1-q) (AAB + BAA )
\]
\end{subequations}
The time is discretized with a small step-size $\epsilon $ (in the code $ \epsilon = 10^{-4}$ for all simulations; faster choices might be possible, but one never should change this parameter for a
set of simulations).
Then  momenta are iterated at half-time units,
starting with a fresh, randomly chosen $
P_{1/2} \in \H_N^2 $.  The momentum $
P_{t+1/2} $ for  $t=1,2,\ldots, \lfloor{ T /\epsilon  }\rfloor $ is obtained from the previous one by letting the force act
during time interval $\epsilon$,\begin{subequations}%
\[
P_{t+1/2} = P_{t -  1/2}   + \epsilon f(A_t,B_t) = P_{ t -  1/2}    - \epsilon N \nabla_X S ( A_t, B_t)
\]
while the matrices $X_{t+1}=(A_{t+1}, B_{t+1})$ are generated by integrating the momentum
$  P_{t+1/2}
= (P_A,P_B)_{t+1/2}$ between time $t$ and $t+1$,
\[
X_{t+1} = X_{t }   + \epsilon  P_{t+1/2}
\]\end{subequations}
This is illustrated in Figure \ref{fig:leapfrog}.
Then let $\tilde X = X_{ \lfloor{ T /\epsilon  }\rfloor}$ and accept or reject according to the rules of  Section \ref{sec:updaterules}.

\subsection{Update rules}\label{sec:updaterules}
First, assume that we want to generate, given a member $X_i \in \H_N^2 $ of the Markov Chain already,
the next element $X_{i+1}$. To this end one proposes $\tilde X \in \H_N^2 $ (Sec. \ref{sec:proposals} told how to propose $\tilde X$).

\begin{itemize}
 \item For a candidate $\tilde X$:
 \[ \text{If } S(\tilde X ) - S(X_i)  < 0 \text{, then accept  } \tilde X = : X_{i+1} \tag{$R_1$} \]
 \item Pick $p$ uniformly distributed from $(0,1)$.
  \[ \text{If }     0 <  S(\tilde X ) -S(X_{i})   <  \frac 1N \log \frac 1p   \text{, then accept  }\tilde X = : X_{i+1}  \tag{$R_2(p)$} \]
  (This, in particular, requires $R_1$ to be false.)
 \item  If  both failed,  let $X_{i+1} =X_i$ and propose another candidate.
\end{itemize}
Figure \ref{fig:R1R2} sketches the previous rules. The first one
tends to spend time at the first local minimum, so without $R_2(p)$ the simulation would remain
there. One thus allows new configurations to climb the potential
but only as far as the increment does not exceed $ N\inv \log (1/p)>0$.


\section{New, dynamic methods in the Python code} \label{sec:Methods}

Let us denote by $\MC(g,h)$ the boolean function,
whose output is decided by convergence (\texttt{True}) or divergence  (\texttt{False})  after performing MCMC on the point  $(g,h)\in \R^2$. Although this function depends of course of the matrix size
and the iteration number $n$, these parameters are kept fixed for each set of simulations, and we
simplify the notation.
By definition, a simulation yields $\MC(g,h)= \False$ if any expectation values diverges at a Markov Chain length $i < n$, i.e. after $i$ iterations  \textit{or} if the hermiticity of the proposed Markov Chain members is irreversibly\footnote{This means,
that after certain fixed number of consecutive attempts to make a Markov Chain
member hermitian failed.} lost at  $i<n$. In that case
the phase diagram marks the point $(g,h) $ with the symbol $\nein$. If all expectation values
converge and hermiticity  is kept during the $n$ iterations, then $\MC(g,h) =\True$ and $(g,h)$ is a denoted by $\ja$.
\\

\begin{figure}
 \begin{subfigure}[t]{0.53\textwidth}\hspace{-.25cm}
\includegraphics[width=.999\textwidth]{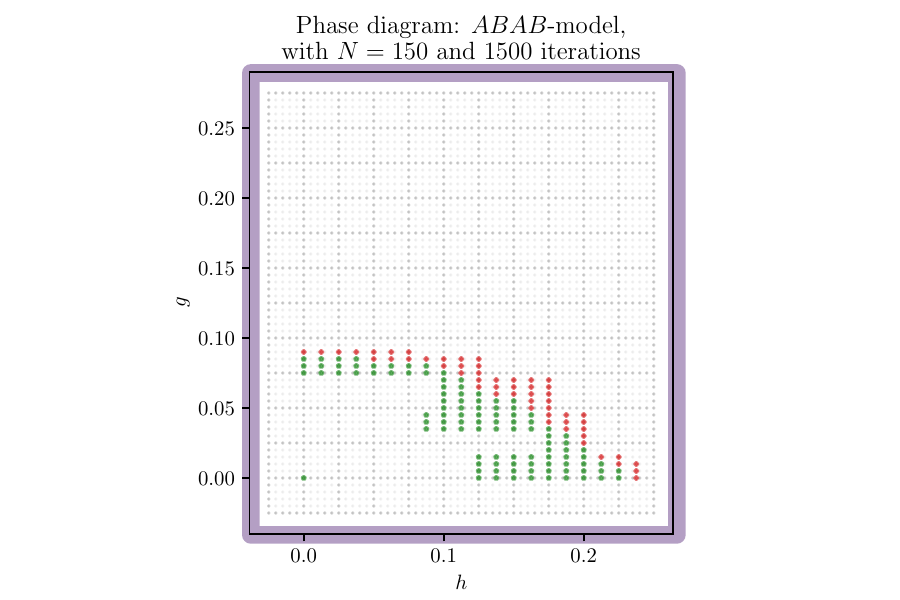}
\caption{\label{fig:naiveA}}
 \end{subfigure}
\hspace{-11ex}\begin{subfigure}[t]{0.53\textwidth}\hspace{-.25cm}
\raisebox{.0\height}{
\includegraphics[width=.999\textwidth]{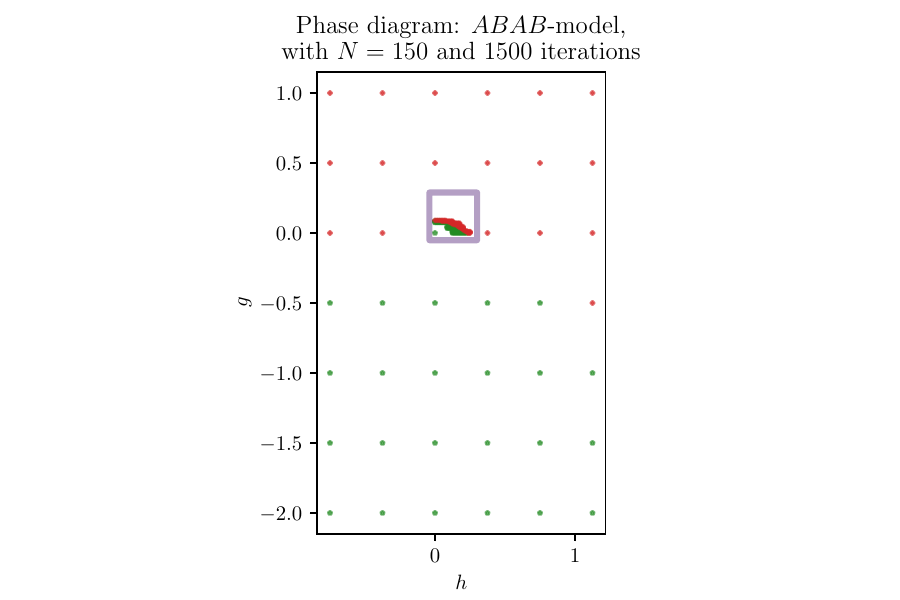}}\qquad \caption{\label{fig:naiveB}} \end{subfigure}
\caption{A naive test of lattice points is not efficient. This leads to  dynamical methods  to search the boundary of the convergent region, see in Sec. \ref{sec:Methods}.\label{fig:naive}}
\end{figure}

 We designed the next routines to dynamically find pairs of opposite truth value that are close enough. This is needed, since a naive strategy based on testing a `static' set (e.g. a lattice) is helpful only initially, in order
to obtain  a glimpse of the phase space structure. However, for more serious experiments
a fixed discrete set turns out to be too expensive. As a matter of fact, we started the simulations
only on a region slightly bigger than that in  Figure
\ref{fig:naiveA}. For this, one can run the script
of \cite{JhaMC} in a double loop; then it was clear, that a dynamical
extension of that code was needed to find the critical curve.
\par

\begin{figure}
\begin{subfigure}{0.99\textwidth}\centering\captionsetup{width=.79\linewidth}%
\includegraphics[width=.327\textwidth]{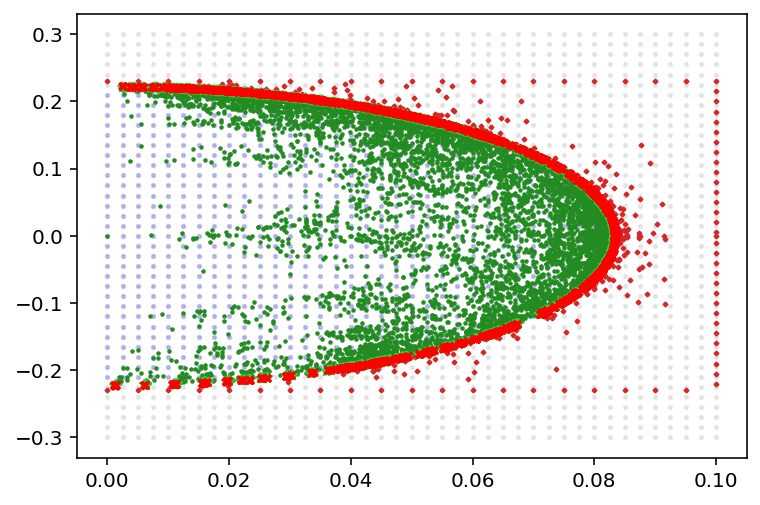}~
\includegraphics[width=.327\textwidth]{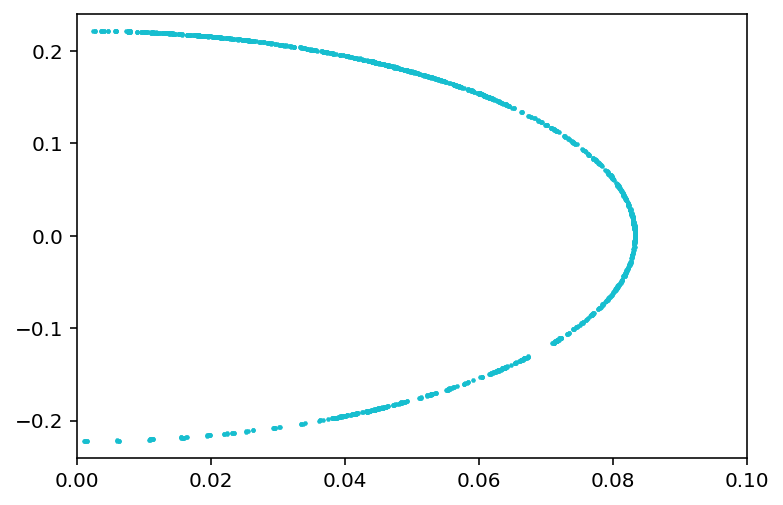}\caption{Testing a
large number of points with a dummy (the characteristic function on an ellipse) to illustrate the midpoint division strategy.
One started here with the outer $\False$-rectangle and a few $\True$-points around the origin.  \label{fig:dummy}}
\end{subfigure}\par\begin{subfigure}{0.99\textwidth}\centering
\includegraphics[width=.475\textwidth]{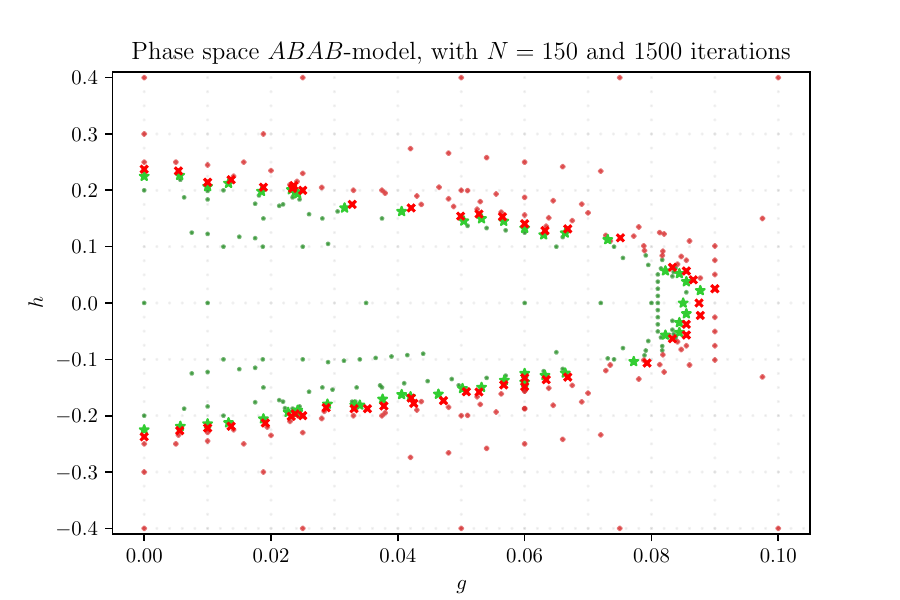}\hspace{-5ex}\captionsetup{width=.79\linewidth}
\includegraphics[width=.475\textwidth]{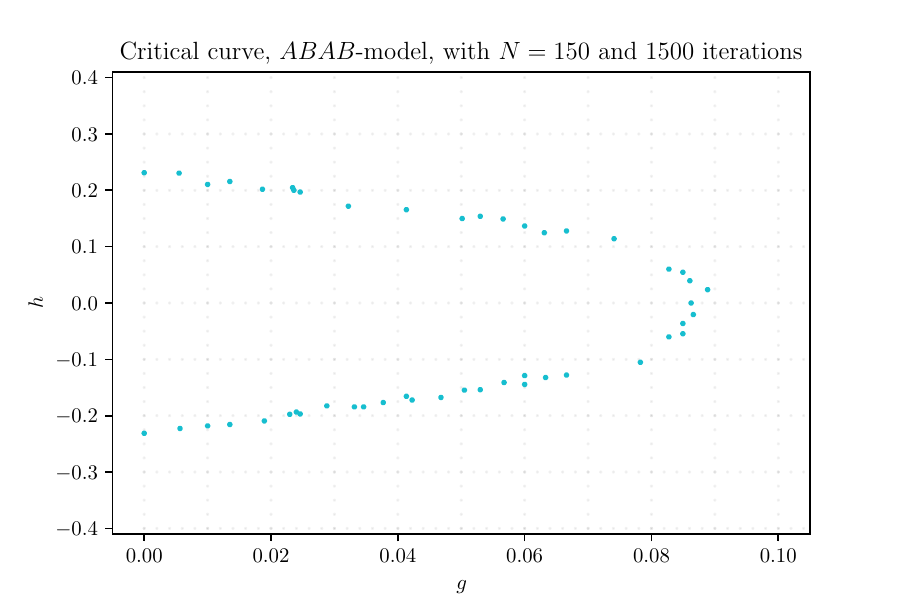}\caption{The \textsc{cpu}-time for $\MC(g,h)$ is  $\approx 400 $ secs., with $N=150$ and $n=1500$. For this small number of interactions $n$,
$\nein$  and each $\ja$ take about the same time, and a dipole $({ \color{limegreen}\star}, \kreuz) $ takes usually at least half an hour.
(This iteration number is not enough to see criticality, and this plot is presented merely with didactic purposes; mind the axis inversion too.) \label{fig:reality}}
\end{subfigure}
\caption{`Expectation (a) vs. reality (b)'. In (a), instead of $\MC$  we use as dummy
a characteristic function. When  points of opposite truth value lie a fixed distance $\delta$ apart (a `dipole')
their middle point is added to the critical curve. \label{fig:expect_vs_reality}}
\end{figure}

Later we will be able to explain in which sense to test a point for criticality is a particularly expensive Monte Carlo integration. This imposes constraints on the number of points to be tested. This section explains the strategies used to thriftily test points,
instead of performing a test like in Figure \ref{fig:dummy}.
Now fix a $\delta>0$. We shall look for \textit{dipoles}, i.e.
two points $(g,h)$ and $(g',h')$ of different truth value
that are at a distance $\delta$ in the $\R^2$ plane. When such a dipole is found,
$\frac12 (g+g',h+h') $ is added to the critical curve and we employ a special notation: ${\color{limegreen}\star}$ and $\kreuz$ instead of $\ja$ and $\nein$, respectively.

\subsection{Midpoint search}\label{sec:midpoint_search}
\begin{figure}[H]
\includegraphics[width=.59\textwidth]{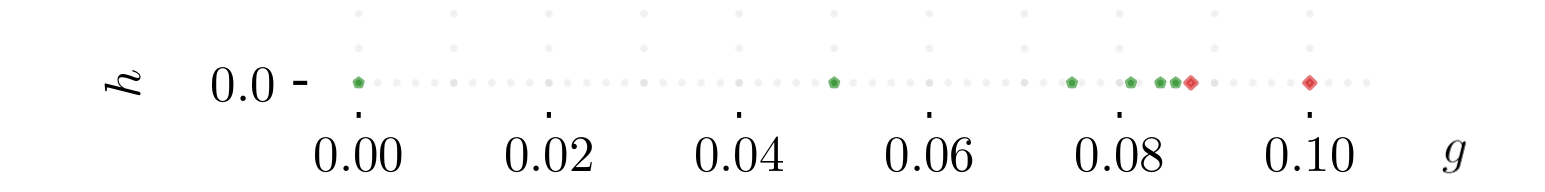}
\caption{Picturing Algorithm \eqref{midpointdiv} with a
simulation at $N=150, n=1500$ at $h=0$ (hence valid for all $q$-models) starting from the
(green) origin and red  point $(0,1/10)$.\label{fig:midpointdiv}}
\end{figure}

Another very simple, but initially useful routine is to use a midpoint search. Start with a pair of points  $ P_0 = (g_0,h_0), Q_0 = (  g'_0,   h'_0) \in \R^2 $ of different truth value ---
w.l.o.g. $\MC(P_0) = \True $.
Suppose that  $(P_i,Q_i) \in  \R^2\times \R^2$ with $\MC (P_i) = \True, \MC(Q_i) = \False$ for some integer $i>0$ were found. If the distance between $P_i$ and $Q_i$
is less than $\delta$, we are done and $(P_i,Q_i)=({\color{limegreen}\star}, \kreuz)$. Else
one  lets
\[ \label{midpointdiv}
(P_{i+1},Q_{i+1}) := \begin{cases}
                     \big( \tfrac{P_i + Q_i }{2}, \,Q_i   \big) & \text{ if }  \MC( \frac{P_i + Q_i }{2})=\True \\
                     \big( P_i, \,  \tfrac{P_i + Q_i }{2}  \big) & \text{ if }  \MC( \frac{P_i + Q_i }{2} )=\False
                    \end{cases}
\]
which by construction is a pair of points of opposite truth value (Fig. \ref{fig:midpointdiv}). The algorithm converges to a dipole, at the latest after $m$
steps, if $2^m  >  (1/\delta)  \sqrt {      (g_0-g_0')^2 + (h_0-h_0')^2  }   $.


\subsection{Radial search}\label{sec:radial_search}
Our Python-code will abandon the evaluation of $\MC(g,h)$   in case of finding divergent quantities or too many failed attempts to make matrices hermitian, and  proposes the next point to be evaluated by $\MC$.
While looking at dipoles, it saves resources to move from potential red points, evaluate these, and if
the result is \False, to move towards the origin
at  a step $\delta>0$ in radial direction, i.e. testing
\[\MC(g-  \delta g/\sqrt{g^2+h^2} ,h-\delta h/\sqrt{g^2+h^2} ) \]
One continues the process
until a green point is found, which must happen since we move towards the origin (however, the code is taking care that one does not unluckily exceeded a maximum of steps; this is tragic, if the initial guess is too far). Since the last red point and the new
first green point are at a distance $\delta$, we obtain a dipole and add it to the critical curve.
The flow diagram for this part of the Python script is depicted in Figure \ref{fig:flowdiag}.
\par

During our experiments we realised that one can save resources by making  $\delta$ dependent on the number of iterations at which the previous point diverged. For instance,
if Python abandons the calculation of $\MC(g,h) $ too soon (say about $n/10$ iterations),
we waste resources by advancing only a step $\delta$.
On the other hand, if $\MC(g,h) $ diverged just before the planned $n$ iterations,
this means that the green point is likely near, and
we loose information by taking the whole step $\delta$. This saves time and was implemented (see App. \ref{sec:Improvements} for this
improvement) but since it complicates the final statistics, so we stuck to
a uniform $\delta = 0.0015$ step to find dipoles in the quadrant $\R_{\geq 0} \times\R_{\geq 0}$, and else we used an angular search described next.

\begin{figure}
\begin{minipage}{.469\textwidth}\centering%
\centering\captionsetup{width=.89\linewidth}%
\hspace{-2ex}
\includegraphics[width=.99\textwidth]{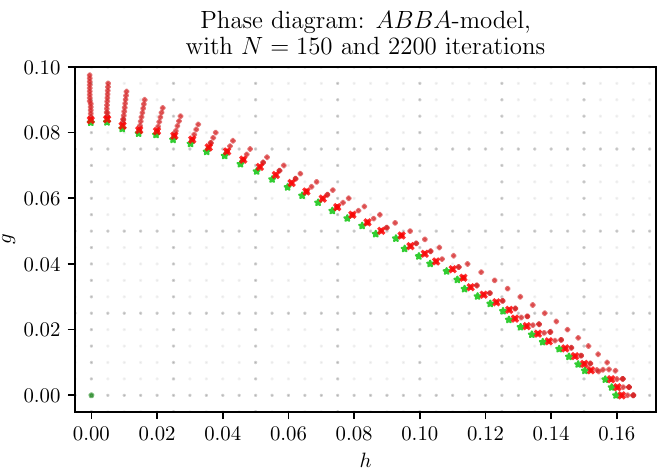}
\caption{\textit{Radial search.} Starting from potentially red points, one advances a distance $\delta$ (here depending on
how soon $\MC$ diverged at the previous point, in the sense of App. \ref{sec:Improvements}) towards the origin until a point along a the same ray yields for $\MC$ a $\True$. The couple of this convergent point and the last divergent point are a dipole denoted by a green star and a red cross.\label{fig:radial_search}}
\end{minipage}%
\hspace{-.41ex}
\begin{minipage}{.52\textwidth}\centering%
\hspace{-.41ex}
\centering\captionsetup{width=.83\linewidth}%
\hspace{-1ex}\includegraphics[width=1.0599\textwidth]{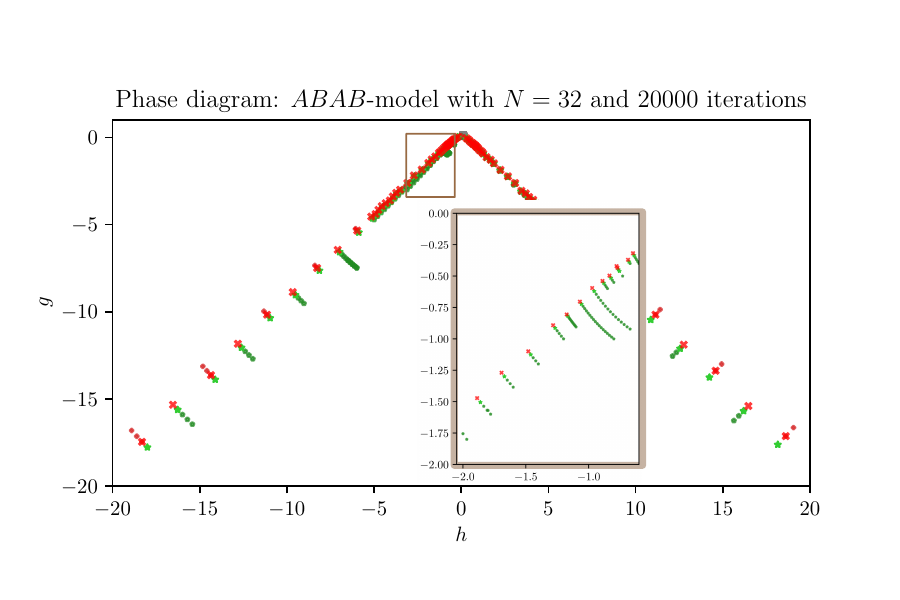}\phantom{AB}\vspace{-3ex}
\caption{\textit{Angular search.} Starting from green points,  one consecutively tests points
on a circumference centered at $ (0,0)$ until a divergent point is found. This is expensive,
e.g. one of the long tails in the zoomed region took $O(10^4)$ secs. ($\approx 3$ hrs.) to find the respective dipole.
A faster, negated version  was also implemented (see the red points with decreasing $|h|$).
\label{fig:angular_search}}
\end{minipage}%
\end{figure}

\begin{figure}
\begin{subfigure}{.49\textwidth}\centering%
\includegraphics[width=.945\textwidth]{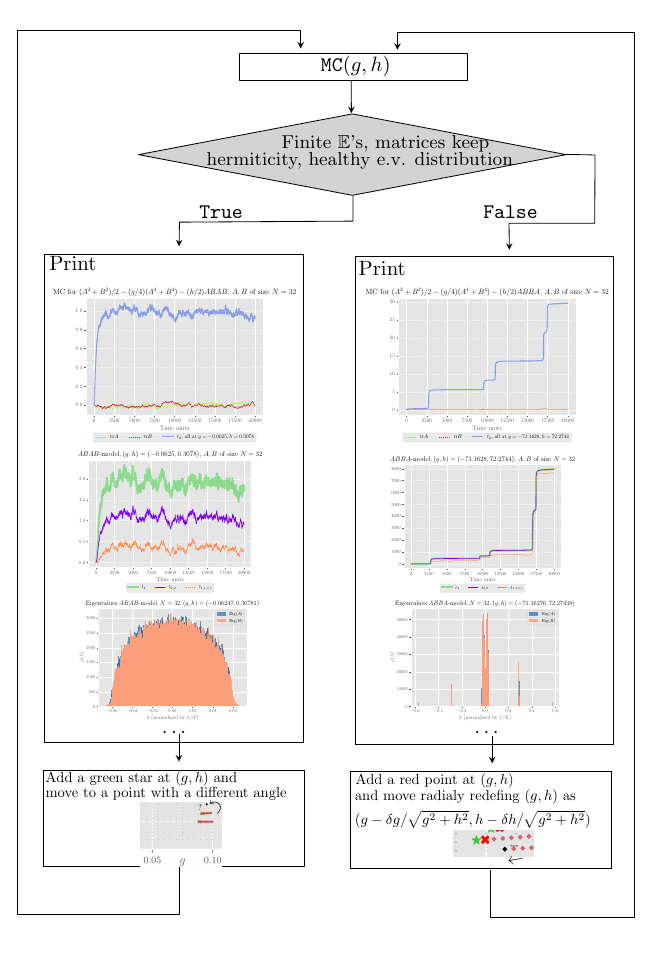}\vspace{-0ex}\caption{Radial search flow diagram.\label{flowA}}\end{subfigure}~%
\begin{subfigure}{.49\textwidth}\centering%
\vspace{.005ex}
\includegraphics[width=.95\textwidth]{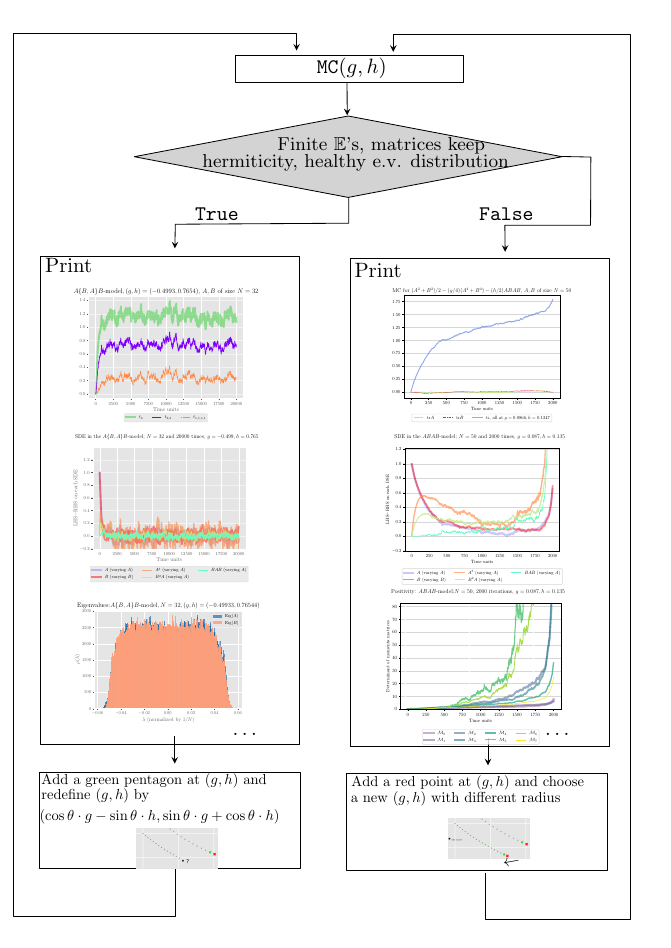}\caption{Flow diagram of the angular search. \label{flowB}}\end{subfigure}
\caption{Flow diagrams for the two main search routines. In the case of (b),
this is the search for  divergent points starting from a convergent point (a negated version is easily derived from this, by swapping $\True$ with $\False$ and green with red points). When the initial points are exhausted,
part of the output is, in each case, a diagram like those in Figs. \ref{fig:radial_search} and \ref{fig:angular_search}.
\label{fig:flowdiag}}

\end{figure}

\subsection{Angular search}\label{sec:angular_search}
Specially to determine the asympotics of each critical curve, a complementary method that tests points along a circumference arc is
helpful. For this, it is again too expensive to stick to our dipole definition in the sense of spatial separation, and we need
a surrogate (keeping the same notation here and in the plots) for angular separation $\alpha$. \par

For $\theta \in [0,2\pi]$, let $ \text{Rot}_\theta (g,h) = (\cos \theta \cdot  g- \sin \theta \cdot h , \sin \theta \cdot g + \cos \theta \cdot h)$
be the point $(g,h)$ rotated by the angle $\theta$.
Starting  with a non-zero red point $Q_0= (g,h) \in \R^{2}-\{\mathbf 0\}$, one
lets $Q_{k+1}=  \text{Rot}_{ k \alpha} (g,h)$ for $k\in \Z_{\geq 0}$ if  $\MC(Q_{k} ) =\False $. Else, we found our  `angular' dipole,
and we draw $\color{limegreen}\star$ for $Q_{k} $ and $\kreuz$ for  $Q_{k-1} $ on the phase diagram. An actual point
(out of two) of the critical curve at radius $  \sqrt {      g^2 + h^2  }$ must be in the arc segment
spanned by these two points (and the other very far).

In particular, since the eigenvalue distribution stresses for large values of $-|g|$ and $-|h|$,
it is convenient not to only trust the previous search for green points. Instead, to be sure that we
find an arc where the critical curve should pass through, we  implemented the `negated' version of
the previous process. Namely, start with a green point now, $P_0 = (g,h)   \in \R^{2}-\{\mathbf 0\}$.
Then let  $P_{k+1}=  \text{Rot}_{ k \alpha} (g,h)$ for $k\in \Z_{\geq 0}$ if  $\MC(P_{k} ) =\True $,
while if  $\MC(P_{k} ) =\False$ then  $({ \color{limegreen}\star}, \kreuz) = (P_{k-1} ,P_{k})$ is a dipole.

For large $n$, we stress that chasing a red point starting with green points takes much longer than chasing
green points starting from a red ones,  since we imposed that Python aborts of the calculation  soon after first clear signs of divergence; since plots are generated also when divergences happened,
we can be sure that we did not abort the calculation too soon (see Fig. \ref{fig:angular_search}).  Another  \textsc{cpu}-time economy aspect is the following: we did not
keep this angular separation constant, so  the maximum of all angular steps is the one that contributes to the
discretisation part of the error bars.

\subsection{Schwinger-Dyson equations  and thermalisation}\label{sec:thermalization}
\begin{figure} \begin{center}\captionsetup{width=.429\linewidth}
\begin{minipage}{.46\textwidth}\vspace{-3ex} \includegraphics[width=.99\textwidth]{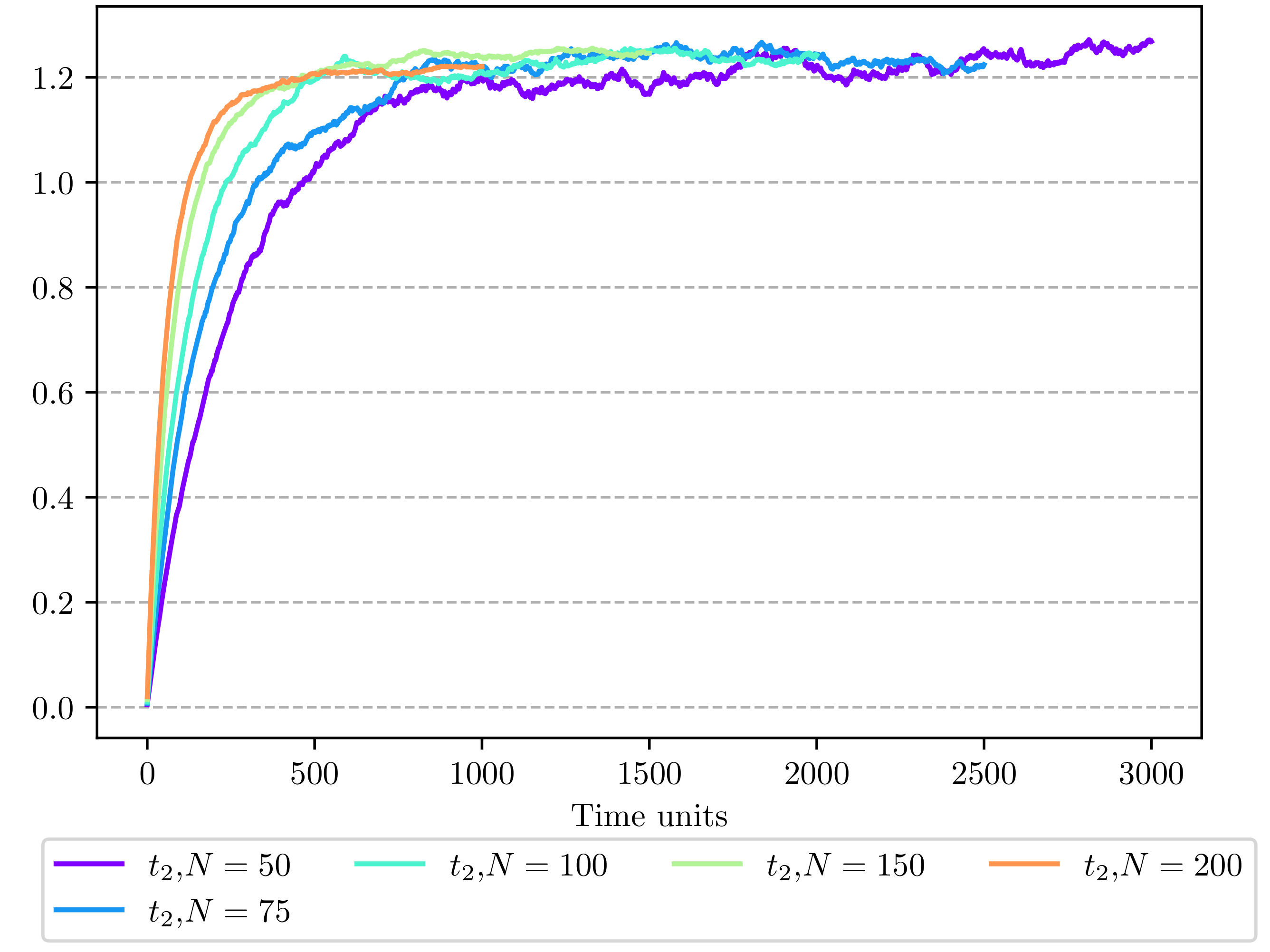}\caption{Thermalisation speed as function of $N$ for the $ABBA$-model at a (convergent) point $(g,h)=(0.05,0.05)$. \label{fig:thermalisation_speed}\small\vspace{5ex}}\end{minipage}\,\,\,\begin{minipage}{.46\textwidth}
\includegraphics[width=.99\textwidth]{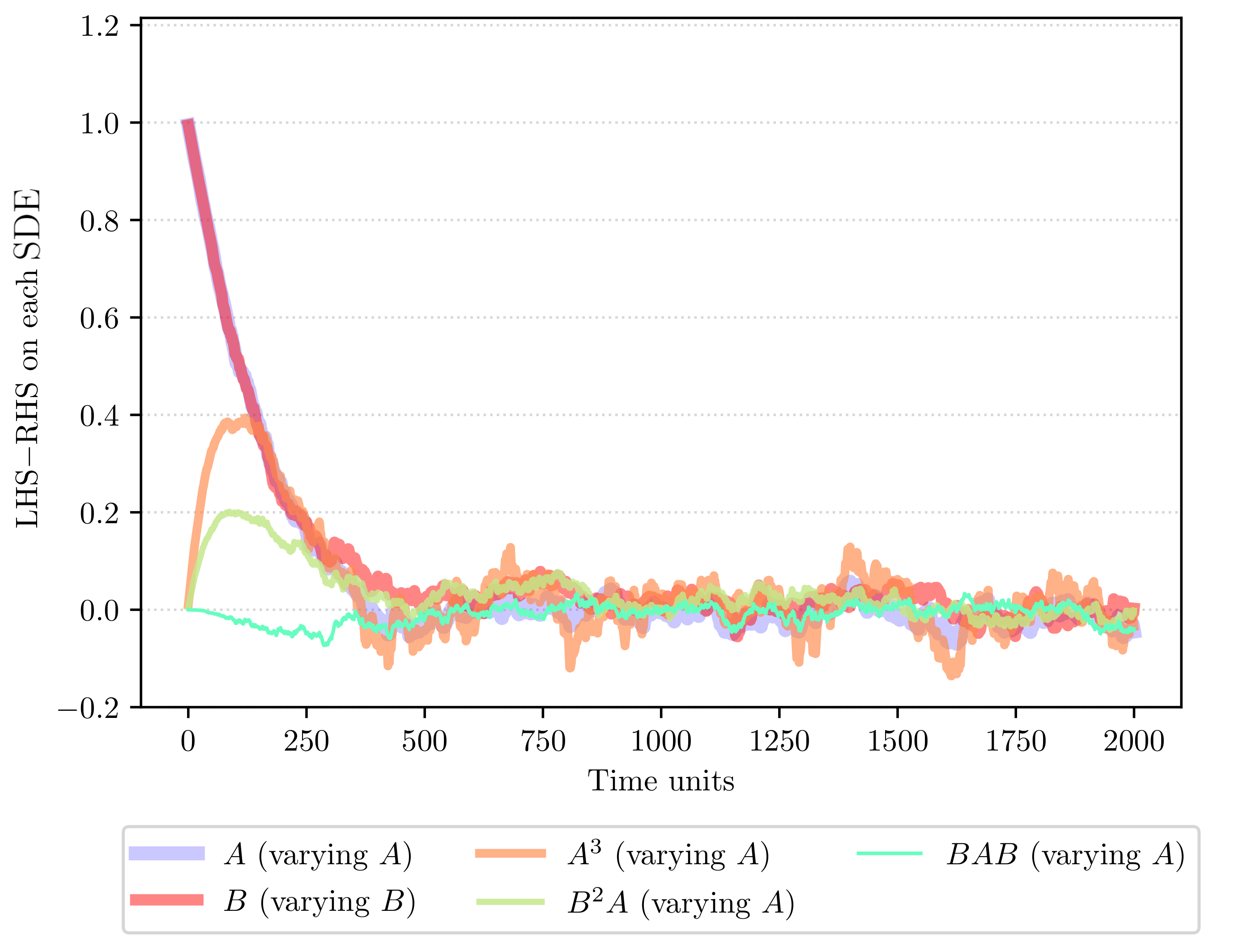}
\caption{Thermalisation and the failure for SDEs to hold. In the plot legend means the five equations tested:  `$W_{A,B}$ (varying $A$)' means that the
field $X=W_{A,B}, Y=0$, while `$W_{A,B}$ (varying $B$)' means  $Y=W_{A,B}, X=0$ in
Table \ref{tab:SDEs}. \label{fig:SDE_thermalisation}}\end{minipage}\end{center}%
\end{figure}

It obvious that the thermalisation speed increases with the matrix size (see Fig. \ref{fig:thermalisation_speed}), but less so is to determine when, for fixed $N$,
thermalisation takes place. Our Python code monitors the Schwinger-Dyson equations during each run.
Each word in $A,B$ or (noncommutative monomial) defines according to Section \eqref{sec:SDEs}
a matrix field, whose respective Schwinger-Dyson equations
are listed in Table \ref{tab:SDEs}. For a fixed word $W_{A,B}$, the random variable
`left hand side (LHS) minus right hand side (RHS)' of the SDE for $W_{A,B}$
is saved and printed as output, as in Figure \ref{fig:SDE_thermalisation}. For future works, this feature can be used as a clearer criterion
for determining the thermalisation time $\tau$ of Section \ref{sec:MCMC}. The rationale is the following: whereas it is hard to choose a criterion to
determine $\tau$ by the stabilization of $t_2,t_{1,1,1,1},t_{2,2}, t_{4},\ldots$
of Eqs. \eqref{moments}  --- partially because it is precisely MC's aim to determine those values --- it is much easier to set a bound $\epsilon>0$ and to determine $\tau$ by looking at the members $\{ (A_i,B_i) \}_{i =1,2,\ldots,n}$ of the Markov Chain and letting $\tau$ be the \textit{first}
index $i_0 \in \{1,\ldots, n\}$ that verifies
\[ \sum_{W_{A,B} }^{\text{finite}}  \Big( \text{LHS-RHS of the SDE for } W_{A,B} \Big)^2\bigg|_{A_{i_0},B_{i_0}}   < \epsilon.\]

\begin{figure}\centering
\begin{minipage}{.46\textwidth}\captionsetup{width=.99\linewidth}
\includegraphics[width=.995\textwidth]{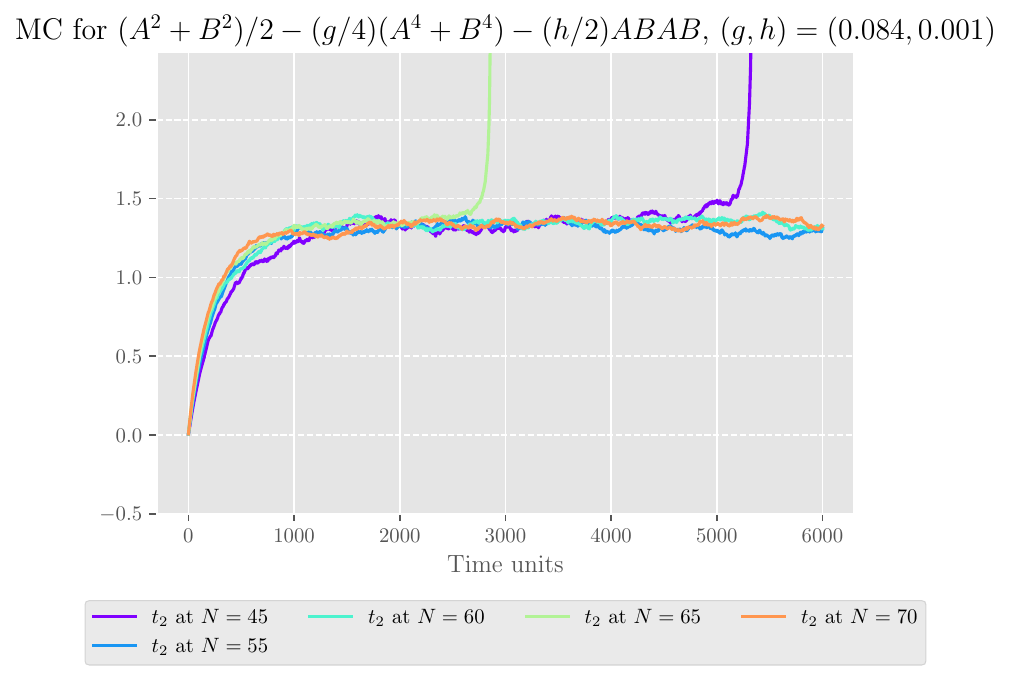}
\caption{This plot explains why we simulate with $n=2\times 10^4$ iterations. To detect criticality, one should be generous with the iterations number to ensure that $\False$-points, like this one, are not only
divergent after the $n$ iterations. \label{fig:severalNs}}\end{minipage}
\,\,\,\,\,\,
\begin{minipage}{.46\textwidth}\captionsetup{width=.79\linewidth}
\includegraphics[width=.995\textwidth]{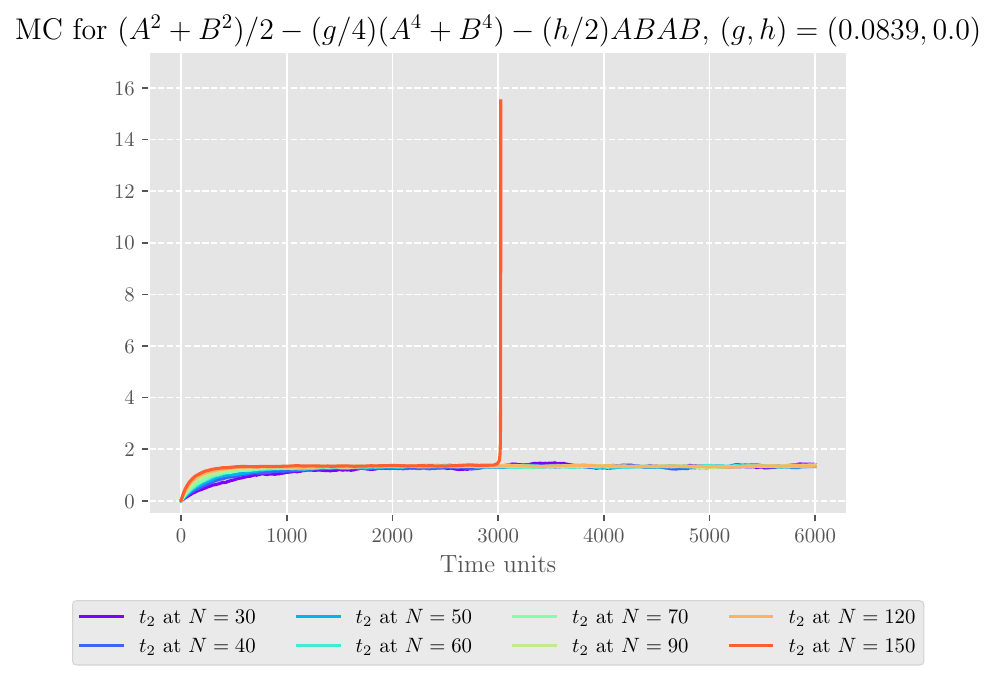}
\caption{This plot shows a MC on a divergent, which `converges' if $n$ is not large enough. Not simulating with a large iterations number requires a   larger-$N$. (See Figs. \ref{fig:severalNs} and \ref{fig:N32N150}.)\label{fig:severalNslarge}}
\end{minipage}
\end{figure}

\section{Results}\label{sec:Results}

After comparing several pairs $(N_1,n_1),(N_2,n_2),\ldots$ of matrix sizes $N_i$ and  and iterations $n_i$,
we chose $N=32$ (see Figs. \ref{fig:severalNs} and \ref{fig:severalNslarge})
as $N=32$  is the smallest value for which comparison with analytically
performed large-$N$ matrix integration would yield  corrections
of the order of $N^{-2} \lesssim 10^{-3}$ (corresponding to  toric and higher-genus corrections $N^{2-2\mathfrak g}$ for $\mathfrak g=1,2,\ldots$).  To compensate this, and in order to ensure that we detect criticality, our Markov chains are quite long ($n=2\times 10^4$). A comparison concerning the stability with respect to matrix size is
presented in Figure \ref{fig:N32N150}.
\\

Now let us finally comment on how to read the results. For all the models,
we approached the critical curve in the quadrant $\R_{\geq 0}^2$ by
using the radial search routine that we developed in Sec. \ref{sec:radial_search}. The plots presented bellow show a   discretisation of the critical curve with error bars, which are meant as follows:  Given an angle $\phi \in  [0,\pi/2]$,
each such plot shows the radius $r=r(\phi)$ at which the critical line is intersected, up to
error bars in the variable $r$ (since $\phi$ is fixed, these errors are truly $1$-dimensional bars, and not boxes). The radial errors (at a fixed angle) are computed according to
\cite{Young:2012kg}; for the radius $r$, $\sigma_r =  [ m(m-1)]^{-1/2}  [ \sum_{a=1}^m (r_a -\bar r)^2 ]^{1/2}$ where $a=1,\ldots,m$ enumerates the experiments, and $\bar r$ is the average of $\{r_a\}_a$. On top of it,
it comes a discretisation error of $\delta$. \par

The main results are in Table \ref{tab:results}. Therein, $\lambda_\pm(q)$ and
$\Theta_\pm(q)$, refer to  (observe the $(h,g)$-axis order, so chosen to ease comparison with \cite{ABAB})
\[\lambda_\pm {(q)} = \text{slope of the $q$-model's critical line when $h\to \pm\infty$}  \label{deflambda}\]
or to the equivalent parameter $  \Theta_\pm {(q)} $ defined as the angle from $h$-semiaxis $\R_{\pm}$ to
said line. Further, one lets $g_0\hp \pm (q)$ be the $g$-coordinate of the intersection point, $(0, g_0\hp\pm (q) )$, of such line with the $g$-axis:

\begin{table}[h!] \scriptsize
\begin{tabular}{ccccc}
 $q$ & Model & Criticality in $\R^2_{\geq 0}$ &   $h\to -\infty$  &  $h\to +\infty$    \\[.3ex] \hline
1 $\vphantom{\displaystyle\int }$& $ABAB$ & Fig. \ref{fig:phase_space_ABAB_c} &
$\lambda_- (1) = +{0.9755} $   &  $\lambda_+ (1)=    -0.9657$    \\[-.95ex]
 & &  & {\tiny $\lambda_- (1)\in [ 0.942,1.010 ]$} &  {\tiny$\lambda_+(1)
 \in  [ -1.000,-  0.93252]$ } \\[1ex]
   & &  &   $\Theta_-(1)   =  ( 44.28 \pm 1) ^\circ  $ &  $\Theta_+(1)  =  ( -44 \pm 1)  ^\circ $
    \\[1ex]  &&& $ g_0\hp -  (1) \approx 0.362  $ & $ g_0\hp  + (1) \approx 0.301$
   \\[2ex]\hline
 1/2 $\vphantom{\displaystyle\int\int}$& $A\{B,A\}A $ & Fig. \ref{fig:phase_space_mix_small_coupling}&
$\lambda_- (\tfrac12) = -0.0023  $   &  $\lambda_+ (\tfrac12)=-0.9764 $    \\[-.5ex]
 & &  & {\tiny $\lambda_- (\tfrac12)\in [  -0.0111, +  0.0064 ]$} &  {\tiny$\lambda_+(\tfrac12)
 \in [-1.0471 ,-0.9105 ]$ } \\[1ex]
   & &  &   $\Theta_-(\tfrac 12) = (-0.135 \pm 0.5 ) ^\circ  $ &  $\Theta_+(\tfrac 12)  =(-44.317 \pm 1)  ^\circ $\ \\[1ex]  &&& $ g_0\hp -  (\tfrac 12 ) \approx   0.167 $ & $ g_0\hp  + (\tfrac 12) \approx  0.490 $  \\[1ex]\hline
0 $\vphantom{\displaystyle\int}$ & $ABBA$ & Fig.  \ref{fig:phase_space_ABBA} &
$\lambda_- (0) = { 0.0027 } $   &  $\lambda_+ (0)= -0.96960 $    \\[-.5ex]
 & &  & {\tiny $\lambda_- (0)\in [ -0.0332, 0.02784 ]$} &  {\tiny$\lambda_+(0)
 \in [-1.0217,  -0.92005]$ } \\[1ex]
   & &  &   $\Theta_-(0)  =( 0.155  \pm 1.75 ) ^\circ  $ &  $\Theta_+(0)= (-44.116 \pm 1.5)  ^\circ $
    \\[1ex]  &&& $ g_0\hp -  (0 ) \approx  0.256 $ & $ g_0\hp  + (0) \approx   0.144$ \\[.4ex]
   \hline
\end{tabular}\normalsize \vspace{1ex}
\caption{Summary of results for the three models. \label{tab:results}}
\end{table}


 \subsection{Bonus: On multimatrix functional renormalisation}\label{sec:bonus}

To the knowledge of the author the only model from the family \eqref{action}
that has been addressed by functional renormalisation is the  $ABAB$-model ($q=1$). In the context of causal dynamical triangulations \cite{Ambjorn:2001br} a phase diagram that resembles to the Kazakov-Zinn-Justin $ABAB$-phase diagram was obtained by \cite{ABAB_FRG} from the
flow given by the following $\beta$-functions:
\begin{minipage}{.54\textwidth}\!\!\!\!\!\!\!\!\raisebox{-4ex}{\includegraphics[width=9.2cm]{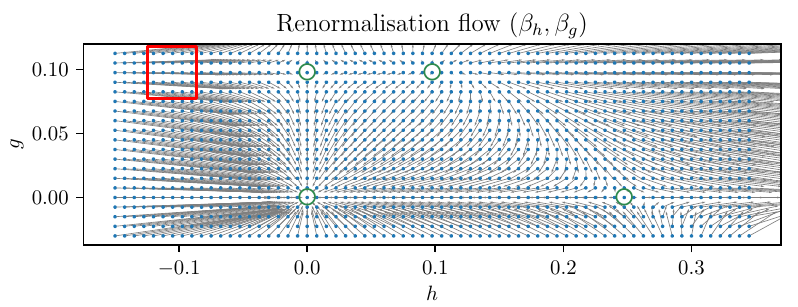}}
\end{minipage}\hspace{-2ex}
\begin{minipage}{.45\textwidth}
\begin{subequations}\label{betaF}%
 \[
 \beta_h &=  [ 1+2 \eta(g) ] h  - F(g)   h^2,  \label{betaFa}\\
 \beta_g &=  [ 1+2 \eta(g) ] g -  F(g)  g^2 ,  \\
 \eta(g)& =\tfrac{8 g}{ 2 g -3 }  ,  \\   F(g) & =  4-\tfrac 45 \eta(g).
 \]\vspace{-1ex}
\end{subequations}%
\end{minipage}\par\noindent%
Here, $\eta(g)$ and $F(g)$ depend on a regulator $r_N$. This  a-priori-functional dependence on $r_N$, luckily boils down to a milder dependence of its moments, $\int r_N^n$, $n=1,2,\ldots $.
Else Eqs. \eqref{betaF} are regulator-independent. A ribbon graph argument  \cite{komentarzABAB}  shows that the Wetterich Equation requires the $h^2$ term  to vanish in Eq. \eqref{betaFa} \textit{independendently} of the chosen regulator
 and truncation (in the number of flowing operators). And yet, the diagram
of \cite{ABAB_FRG} does  --- more or less after axes rescaling $(h,g)\to  (9 h/10, 10 g/12)$  --- look like
the Kazakov--Zinn-Justin phase diagram, and even has similar fixed-points (above encircled). Why? \par

\begin{figure}\captionsetup{width=.99\linewidth}
\centering%
\begin{subfigure}{.486\textwidth}\centering
\!\!\!\!\!\includegraphics[width=.999\textwidth]{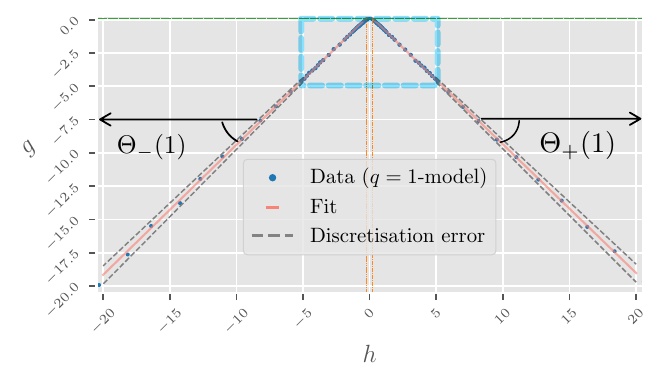}\caption{Critical curve and bounds for the large coupling asymptotics, with angles $\Theta_\pm(1)$ and their  error represented with the dashed gray lines.
The dashed fat line rectangle is zoomed in Fig. \ref{fig:phase_space_ABAB_b}. \label{fig:phase_space_ABAB_a}}
\end{subfigure}\hspace{-1ex}%
\begin{subfigure}{.486\textwidth}\centering\vspace{-7ex}%
\includegraphics[width=1.01999\textwidth]{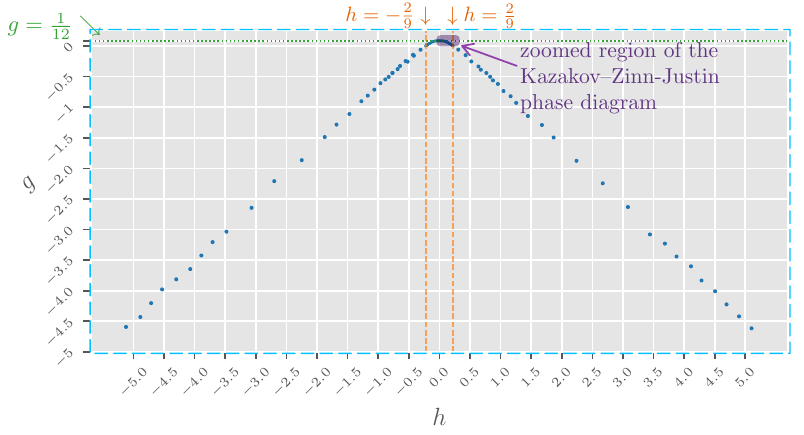}
\caption{Zoom of the dashed region in Fig. \ref{fig:phase_space_ABAB_a}.\label{fig:phase_space_ABAB_b}}\end{subfigure}\newline\noindent
\begin{subfigure}{.9\textwidth}\centering%
\includegraphics[width=.783\textwidth]{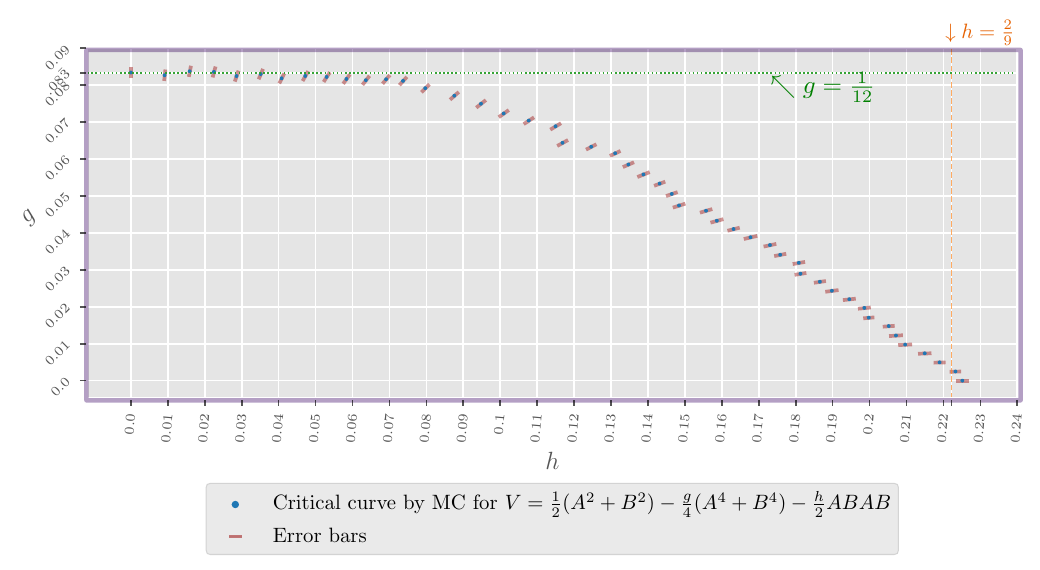}
\caption{Zoom of the shaded region in Fig. \ref{fig:phase_space_ABAB_b}.
This plot is to be compared with \cite[Fig. 4]{ABAB}. Error bars are 1-dimensional and radial (see Sec. \ref{sec:Results}).\label{fig:phase_space_ABAB_c}}
\end{subfigure}
\begin{subfigure}{.9\textwidth}\centering
\quad \includegraphics[width=.83\textwidth]{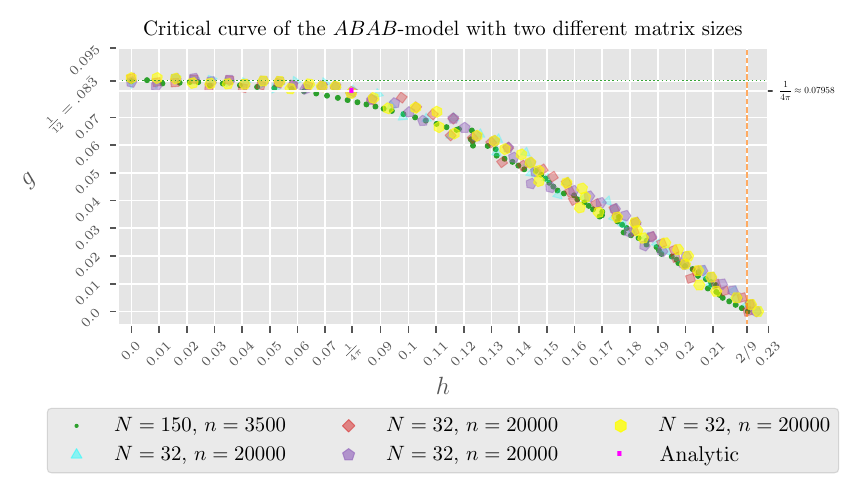}\caption{Output of some  of the experiments among which only one with different matrix size and number of iterations ( in the final statistics we consider only those with $n=2\times 10^4$). `Analytic' for the magenta pixel at $(g,h)=(1/4\pi,1/4\pi)$ is the
Kazakov--Zinn-Justin  critical point. It is special in the sense of two branch cuts in the space of maximal weights of the character expansion merging there. This is hit by all experiment shown, except that at $N=150$ (a higher number of iterations would push it outwards, where it should be).  \label{fig:N32N150}}
\end{subfigure}%
\caption{Phase diagram of the $ABAB$ or $q=1$ model.
Here, all simulations were performed at $N=32$ and $n=2\times 10^4$, except in (d), in order to compare  matrix sizes.
\label{fig:phase_space_ABAB}}
\end{figure}

\begin{figure}%

\begin{subfigure}{.95\textwidth}\centering%
\captionsetup{width=.86\linewidth}\centering\captionsetup{width=.80\linewidth}
\includegraphics[width=.87\textwidth]{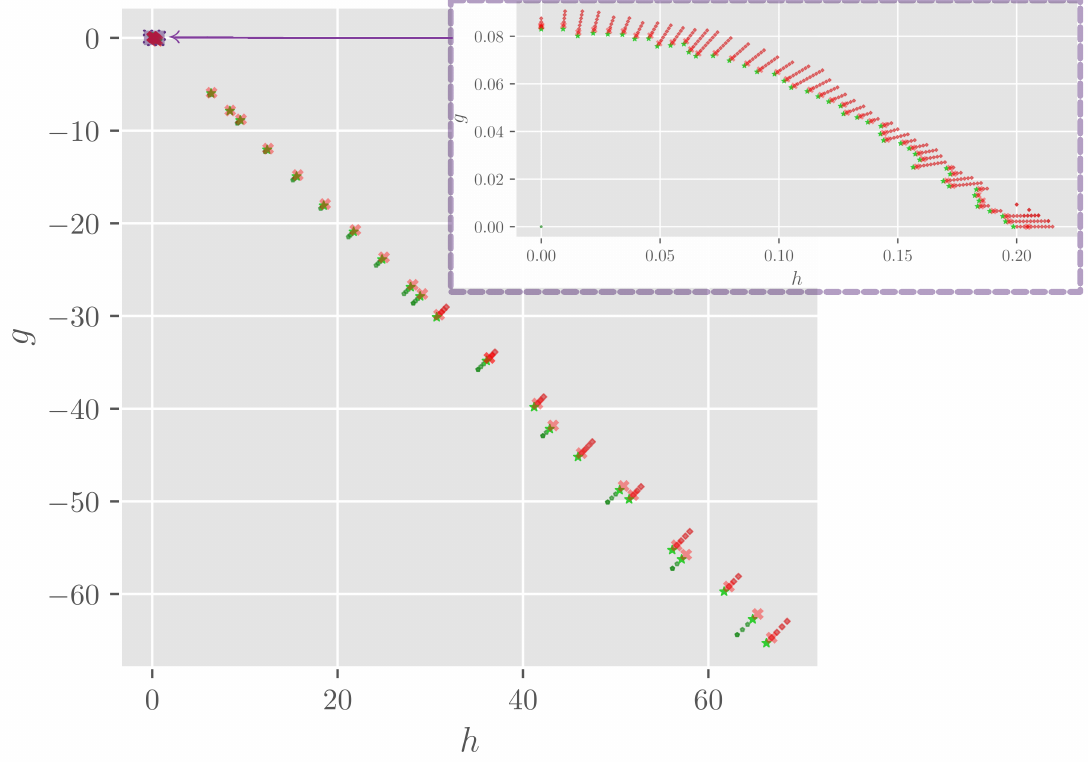}\caption{Large $h$ (or large $-g$) asymptotics of the phase diagram of the $q=\tfrac 12$-model. \label{fig:mix_gnegative}}%
\end{subfigure}


\begin{subfigure}{.69\textwidth}\centering%
\captionsetup{width=.97\linewidth}!\!\!\!
\!\!\!\!\includegraphics[width=.997\textwidth]{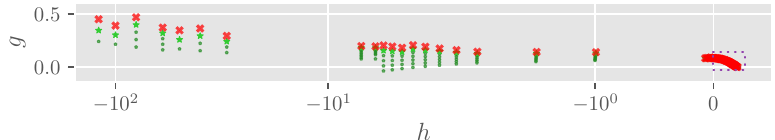}
\vspace{.009ex}\caption{Large $-h$ asymptotics  of the phase diagram of the $q=\tfrac 12$-model (observe the negative log-scale of  $h$ and the ordinary one of  $g$).\label{fig:mix_hnegative}}%
\end{subfigure}

\vspace{1ex}

\begin{subfigure}{.55\textwidth}\centering%
\captionsetup{width=.7\linewidth}
\hspace{-2ex}\includegraphics[width=.9963\textwidth]{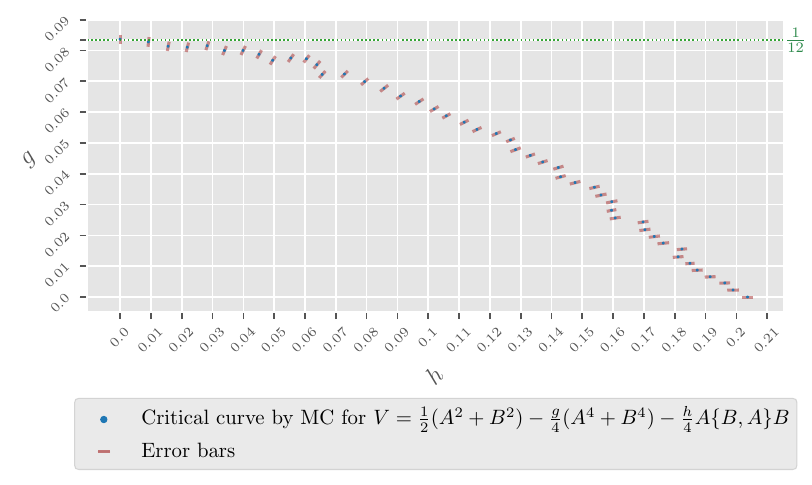}
\caption{Segment of the critical curve for positive (hence small) couplings
of the $A\{B,A\}B$-model. The dashed line is $g=1/12$. \label{fig:phase_space_mix_small_coupling}}
\end{subfigure}\hspace{-2ex}
\begin{subfigure}{.44\textwidth}\centering\vspace{-2ex}%
\includegraphics[width=1.09963\textwidth]{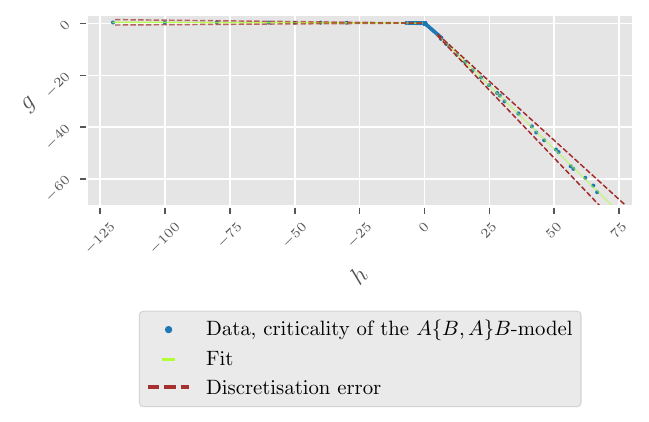}\caption{Large coupling asymptotics.\label{fig:phase_space_mix}}
\end{subfigure}
\caption{Phase diagram of the $A\{B,A\}B$-model in its several scales. \label{fig:phase_space_mix}}
\end{figure}

The present answer relies on two observations.
First,  the
critical segments of the $ABAB$-model
and of the $A\{B,A\}B$-model in the positive couplings quadrant (see Figs.
\ref{fig:phase_space_ABAB_b} and
\ref{fig:phase_space_mix_small_coupling}) are more similar among themselves
than the renormalisation phase portrait is to any of these two; further, if one scales down Diagram \eqref{betaF} by a homogeneous 1/12 factor, the fixed points on the axes match those of the  $A\{B,A\}B$-model, not the $ABAB$-model's.  The second and decisive point is  Figure \ref{fig:phase_space_ABAB}
showing the $h \mapsto -h$ symmetry of the $ABAB$ critical curve.  Were the set of Eqs. \eqref{betaF} the $\beta$-functions of the $ABAB$-model,
then the $h\leq 0 $ region would be the specular image of the $h\geq 0$ region; this is not the case,
in particular because  the specular image $ (-0.1,0.1)$ of the encircled fixed point $ (+0.1,0.1)$ has
no fixed point near\footnote{In the phase portrait of \eqref{betaF} fixed points are those without emerging arrows.} (see the square box). Instead, the flow line through $ (-0.1,0.1)$ extends from $-\infty$
parallel to the $h$-axis to $(0,0.1)$.
The parallel flow lines of \eqref{betaF} at $h=-\infty$ share the critical straight slope $\lambda_-(\tfrac 12) = 0 $
up to the bounds of Table \ref{tab:results}. The significant difference with $\lambda_-(1) = 1 $ is decisive to tell apart the $h\to - \infty$ asympotics of both models. Thus the ribbon-graph proof in \cite{komentarzABAB}
that prohibits the $h^2$ term in Eq. \eqref{betaFa} is not empty formalism --- the present Monte Carlo simulations  confirm that the presence of $h^2$ in Eq. \eqref{betaFa} impacts the $ABAB$-model's flow and kicks it towards a $q$-model's flow with $q\leq \tfrac 12$ (whose $\beta_h$-function, without it
being exactly Eq. \eqref{betaFa},
does accept a $h^2$ term in its 1-loop structure). \par
There exists another approach to functional renormalisation of multimatrix models started in
 \cite{Perez-Sanchez:2020kgq}, which will likely struggle to find the
 multimatrix model critical behaviour [it reports some critical constants like the
 present $1/4\pi$, but a more interesting result would be to obtain this after with a
regulator solves an (integro-differential) equation].
The author wrote that article with a random geometry problem in mind,  the only renormalisation-paper being its  spin-off \cite{Perez-Sanchez:2021mvi}.
 Both work under the commonly accepted assumption that the renormalisation flow can be computed in terms of  unitary-invariant operators (products of pure traces of words), and removing it might led to progress.
The `noncommutativity' of the differential operators of \cite{Perez-Sanchez:2021mvi} is just language
that makes it easier to implement code, but which only rephrases the fact that
in general two $A, B \in \H_N$ will not commute. Thus, in the present context, `noncommutativity' is just a counterpart to
the appearence of diagonal matrices to compute the flow as in \cite[p. 13]{ABAB_FRG}.

\begin{figure}
\begin{subfigure}{.55\textwidth}\centering%
\captionsetup{width=.7\linewidth}\includegraphics[width=.93\textwidth]{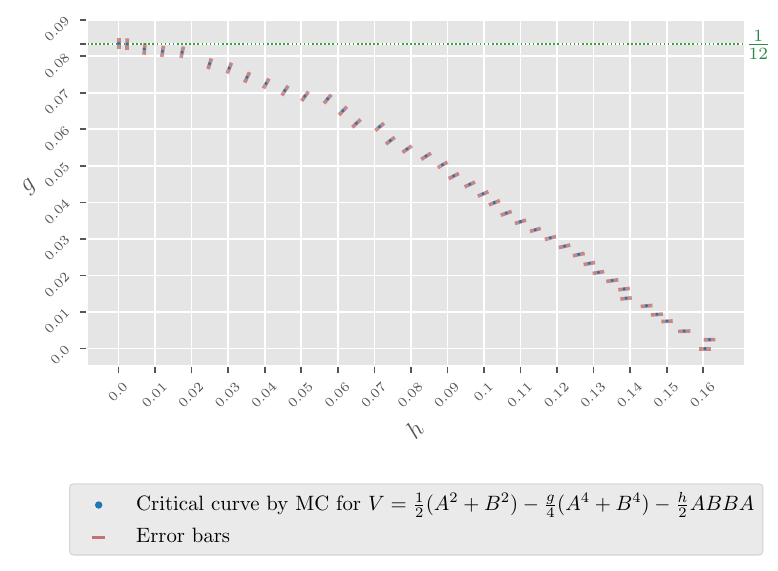}
\caption{Segment of the critical curve of the $ABBA$-model
in the $\R_{\geq 0}^2$-quadrant. \label{fig:phase_space_ABBA_smallcoupl}}
\end{subfigure}\hspace{-2ex}
\begin{subfigure}{.44\textwidth}%
\captionsetup{width=.86\linewidth}\includegraphics[width=1.09963\textwidth]{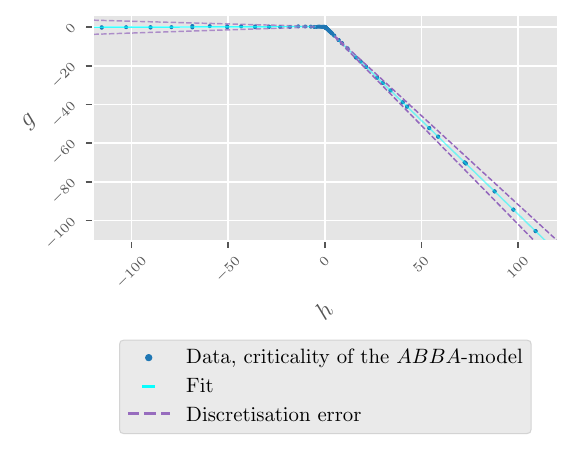}
\caption{Large coupling behaiviour of the critical curve of the $ABBA$-model.\label{fig:phase_space_ABBA_largecoupl}}
\end{subfigure}
\caption{Results for the $ABBA$-model ($q=0$).\label{fig:phase_space_ABBA}}
\end{figure}

\begin{figure}
\begin{minipage}{.6\textwidth}\centering\captionsetup{width=.90\linewidth}
\includegraphics[width=.85\textwidth]{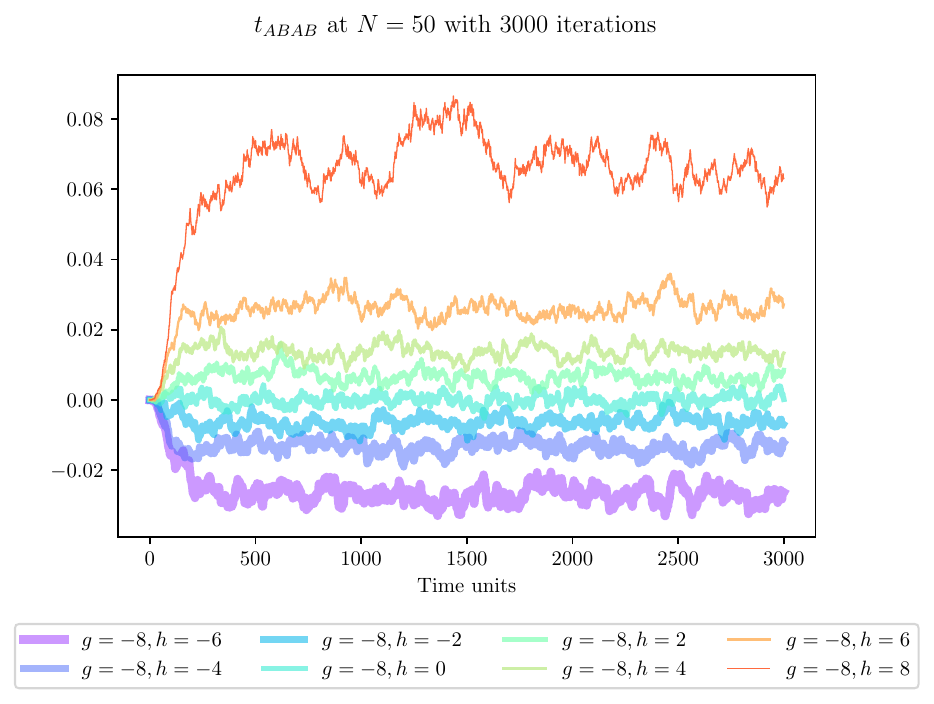}
\caption{Skew symmetry of the expectation of the $\Tr ABAB$ operator in the $ABAB$-model,  $\mathbb E_{g,h} \hp {q=1}  [ \Tr ABAB] = -\mathbb E_{g, - h} \hp {q=1} [ \Tr ABAB] $ \label{fig:sign_ABAB} is illustrated for $ h= 0, \pm 2,\pm 4,\pm 6$ at fixed $g=8$.
(notice the negative of $h=8$ was not plotted, whence the aparent lack of symmetry).
}
\end{minipage}\hspace{-3ex}~
\begin{minipage}{.4\textwidth\captionsetup{width=.36\linewidth}}\hspace{-2ex}\includegraphics[width=1.05\textwidth]{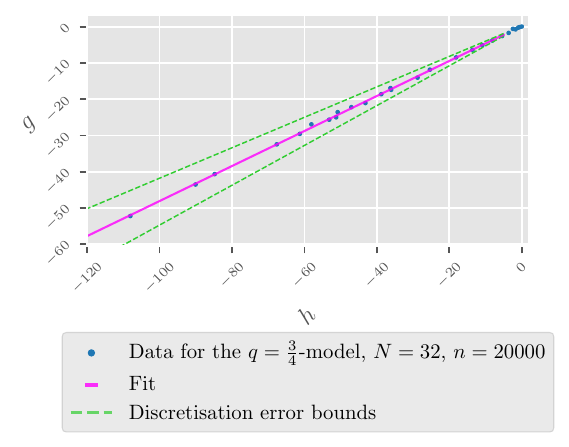}
\caption{The $h\to -\infty$ asymptotics of the critical curve for $q=\tfrac 34$.
The divergent part is above the magenta (solid) line. \label{fig:q0.75}}
\end{minipage}
\end{figure}

\section{Conclusions} \label{sec:conclusions}

We determined the critical curves of three members of the family
of models with interaction $ - \frac{g}{4}  \Tr (A^4+B^4)
- \frac{ h}{2}  \Tr  [ q A BA B + (1-q) ABBA ]$, $q \in [0,1]$
in the positive quadrant, and the slopes of their asymptotics.
A closer relative is another family of two-matrix models,  expressed in \cite{Kazakov:2021lel}  in terms of  a $\mathfrak q$-deformed commutator with an interaction $ \Tr ( [A,B]_{\mathfrak q} [A,B]_{\mathfrak q})$  that is reminiscent of fuzzy geometries \cite{OConnor:2013gfj,Steinacker:2026qzk,BarrettGlaser,Perez-Sanchez:2019tuk,hessam2022bootstrapping,Perez-Sanchez:2021vpf,DArcangelo:2026kka}.  To the knowledge of the author that $q$-deformed commutator model has not been studied.
\par

Having obtained very similar phase spaces for
$q=0$ and $q= \tfrac 12$
there is likely a desert in between, while the situation for $q\geq \tfrac 12$ promises
diversity.  In fact, the short exploration of the $q=\tfrac 34$-model yields, in that limit,
a slope of $\lambda_-(\tfrac 34) \approx 0.48 \pm 0.066 $ (corresponding to
an an angle $\Theta_- (\tfrac 34) $ between  $22.8^\circ$ and $28.8^\circ$ formed with the negative $h$-axis) as Figure \ref{fig:q0.75} shows.\par

Since the $q=1$ model has an interpretation in terms of Causal Dynamical Triangulations
\cite{Ambjorn:2001br, ABAB_FRG}. If the $ABBA$-term can also be interpreted as gluings,
we believe that the present results could be useful there.
Finally, Models \eqref{action} can be simultaneously
studied from viewpoints like positivity bootstraps or the renormalisation group.
These and analytic approaches motivate are the subject of the next part(s) of this article.

\small
\appendix

\section{Further improvements in the algorithm}\label{sec:Improvements}
\noindent
\begin{minipage}{.659\textwidth}
Here we go back to the context of  Section \ref{sec:Methods},
in which a strategy to
economise simulation time was promised.
Assume that we start at a red point $(g,h)$, $\MC(g,h) =\False$.
This means that the planned $n$ iterations were not completed,
and let $\mathfrak a$ be the fraction of those missing (say,
in percent).  Having started our experiments
with a uniform step $\delta$, we did not use this feature to generate
data, but we
remark that a dependence of the step $ \delta(\mathfrak a) $
on $\mathfrak a$, like the one shown in the right was useful.
\end{minipage}
~~~
\begin{minipage}{.283\textwidth}
\includegraphics[width=.95\textwidth]{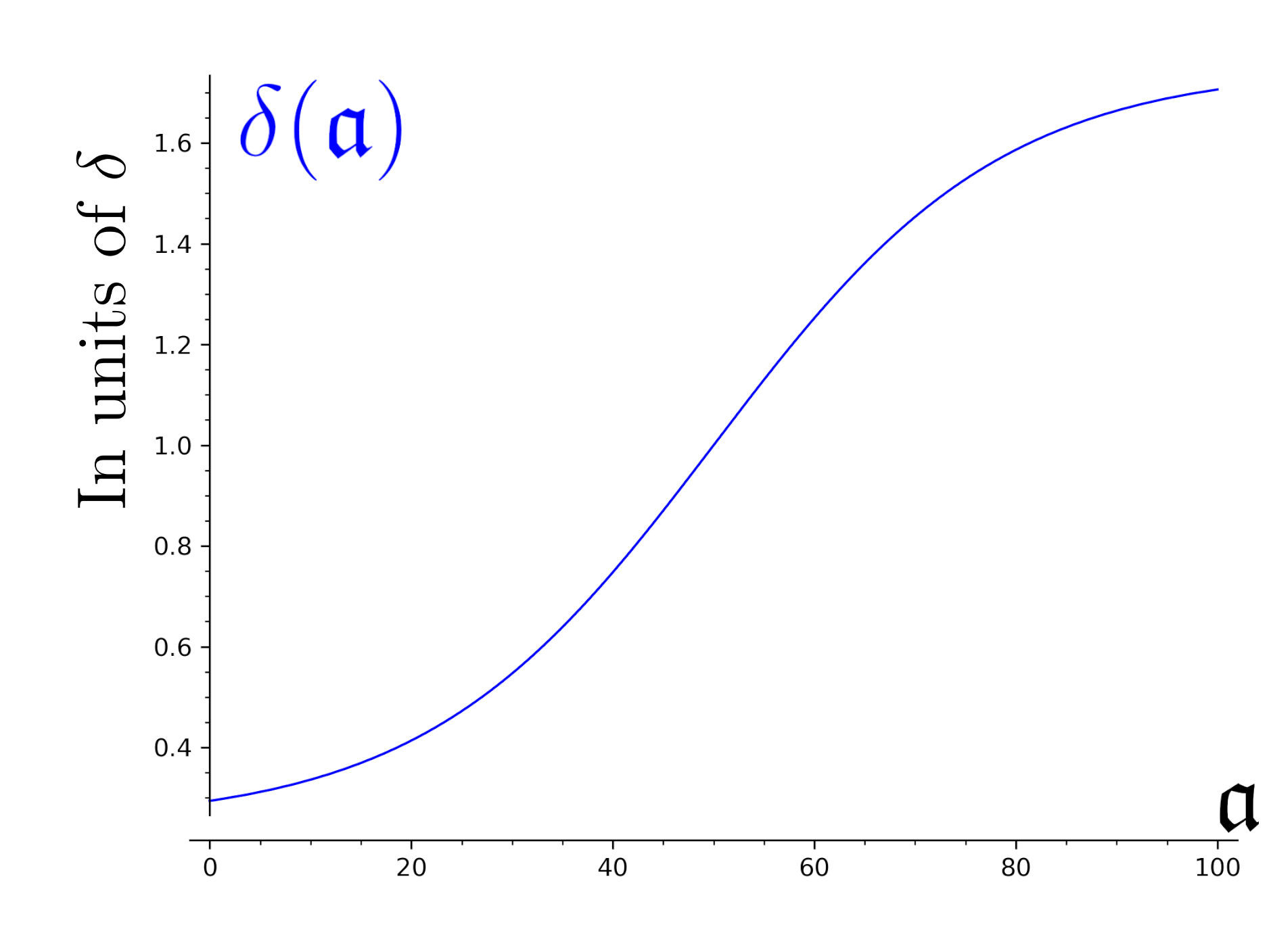}
\end{minipage}
\\

This means that in the
radial search of Section \ref{sec:radial_search}, the next point to
be tested after the divergent point $(g,h)$ is
\[\MC\bigg( g-\delta(\mathfrak a)  \frac{ g}{\sqrt{g^2+h^2}} ,h-\delta(\mathfrak a)  \frac{ h}{\sqrt{g^2+h^2}} \bigg). \]
A second improvement is on top of it is still possible.
From Ineqs. \eqref{hierarchies_examples} one has $ \frac{1}{2}| \Tr ABAB |  \leq
 \frac{1}{2} \Tr ABBA   \leq  \frac 14 \Tr (A^4 +B^4) $
which means that, for any $q$,  Models \eqref{action} are more sensitive to changes to $g$ than those in $h$.
In order to adjust the measurement resolution in each direction
we implemented for future simulations a
step
$\delta(\mathfrak a)  $
that depends on $\phi$ of the point $(g,h)$ measured from the $g$-axis. In summary, we start with a red point $(g_0,h_0)\in \R_{\geq 0}^2$, let $\mathfrak a_0:=1/2$ by definition, and iterate as follows:
as far as $
\MC\big(g_{k},h_{k} \big)  =
\False$ and Python abandoned the simulation leaving
uncalculated a fraction $\mathfrak a_{k}$ of the planned iterations,
we evaluate  $\MC\big(g_{k+1},h_{k+1} \big)$ where
\[
(g_{k+1},h_{k+1} )  = \big(  g_{k}-\delta(\mathfrak a_{k} ,\phi)
\cos(\phi) ,h_k-\delta(\mathfrak a_{k} ,\phi)  \sin \phi \big).
 \]
It is trivial to verify that dropping the subindex $k$ for $\phi_k$ (the angle of $(g_k,h_k)$ with the $g$-axis) is legal, as we stay on a ray
with the same angle $\phi$. Also we move towards the origin, since we started in the  quadrant
$ \R_{\geq 0}^2$, and at the latest $\MC(0,0) = \True$ trivially. However, for practicality we set a cut off for $k$, after which we start with a new ray, i.e. with a $ (g_0',h_0')  $ with  different $\phi'$.

\section{Notation}\label{sec:Glossary}
     \thispagestyle{empty}
\fontsize{10.5}{13.5}\selectfont
 \begin{tabularx}{.85\textwidth}{lX}
$*$ & if $M$ is a matrix, $M^*$ is its transpose and complex conjugate (elsewhere $M^\dagger$)\\
$\bar z, \bar M_{a,b} $ & complex conjugate of $z$, $M_{a,b}$\\
\ja  & point $(g,h)$ at which $\MC(g,h) = \True$, or just green point\\
\nein  & point $(g,h)$ at which $\MC(g,h) = \False$, or just red point\\
$\color{limegreen}\star$ &   point $(g,h)$ at which $\MC(g,h) = \True$ in a $\delta$-neighbourhood or angular neighbourhood of a
red point (then marked \kreuz ) \\
\kreuz  &   point $(g,h)$ at which $\MC(g,h) = \False$ in a $\delta$-neighbourhood  or angular neighbourhood  of a
green point (then marked ${\color{limegreen}\star}$)  \\
 $A,B$ & two hermitian random matrices of size $N$ \\
  $\{A,B\}_q $&  $ qAB + (1-q)BA$  \\
    $\{A,B\}  $&  $AB+BA$ \\ `dipole' & pair $({\color{limegreen}\star}, \kreuz)$   \\ $\delta $ & dipole separation (in our experiments always $\delta = 0.0015$) \\ $\E$  & expectation value, written in full: $\mathbb E \hp q_{g,h; N}$ \\
   $f=(f_A,f_B)$ & force $f=\nabla_X S$, with $\nabla_X=(\partial_A,\partial_B)$ the matrix gradient    \\
 $(g,h)$ & coupling constants \\
 $\H_N$ & space of hermitian matrices of size $N$ \\
 $\MC(g,h)$ & boolean function of $(g,h)$,  $\True$ if integrals converge
 after $n$ iterations (in our results always $n=2\times 10^4$) \\
 $\lambda_\pm (q)$ & slope of the $q$-model's critical curve at $h\to \pm \infty$\\
 $N$ & always the matrix size \\
 $n$  & always the number of iterations (length of the Markov Chain) \\
 $q$ & parametrises the convex combination $ (1-q) ABBA+q ABAB$ \\
$\Theta_\pm (q) $ & $\measuredangle( \pm h$-axis,  $q$-model's critical line  at $h\to \pm \infty$) \\
 $S$ or $S_{g,h} \hp q(A,B)$ &  $ \frac{1}{2} \Tr (A^2+B^2) - \frac{g}{4}  \Tr (A^4+B^4)
- \frac{h}{2}  \Tr (  A \{B,A\}_q B)  $\\
SDE & Schwinger-Dyson or Dyson-Schwinger or loop equations \\ \end{tabularx}

   \begin{tabularx}{.85\textwidth}{lX}
$\Tr A_1A_2\cdots A_l $ & brackets economy meaning $\Tr (A_1A_2\cdots A_l) $ and not $ ( \Tr A_1) A_2\cdots A_l$ \\
$t_2$  &  $ ( 1/{2N})\E [ \Tr (A^2 + B^2)  ]$ \\
$t_4$  &  $  ( 1/{2N})\E [ \Tr (A^4 + B^4)  ]$ \\
$t_{2,2}$  &  $ ( 1/{N}) \E [ \Tr (ABBA)  ]$ \\
$t_{1,1,1,1}$  &  $ ( 1/{2N})\E [ \Tr (ABAB)  ]$\\
$\tau$ & thermalisation time \\ $V$ & $\Tr V = S $ \\
$X, X_i,\tilde X$ & respectively: $(A,B) \in \H_N^2$; a Markov Chain member $(A_i,B_i) \in \H_N^2$;  or a proposal $(\tilde A,\tilde B) \in H_N^2$ for the Markov Chain \\
$Z$ or $Z\hp{q} _N (g,h) $ & $  \int \ee^ { - N S_{g,h} \hp q(A,B) } \dif A\,\dif B $ \\ \end{tabularx}
  \hspace{-.02cm}
     \fontsize{11.6}{15.0}\selectfont
 \newcommand{\etalchar}[1]{$^{#1}$}
\bibliographystyle{alpha}

\end{document}